\documentclass[5p,times,twopage,twocolumn,a4paper]{elsarticle}
\usepackage{graphicx}
\usepackage{multirow}
\usepackage{amsmath,amssymb,amsfonts}
\usepackage{amsthm}
\usepackage{mathrsfs}
\usepackage[title]{appendix}
\usepackage{xcolor}
\usepackage{textcomp}
\usepackage{manyfoot}
\usepackage{booktabs}
\usepackage{algorithm}
\usepackage{listings}
\usepackage{threeparttable}
\usepackage[hyphens]{url}
\usepackage{babel}
\usepackage{adjustbox}
\usepackage{soul}
\usepackage{lscape}
\usepackage[cachedir=.]{minted}
\usepackage{pbalance}   % <<-- balance references on last page
\usepackage{orcidlink}

\begin{document}
\begin{frontmatter}
\journal{arXiv}
\title{PELLI: Framework to effectively integrate LLMs for quality software generation}

\author{Rasmus Krebs, Somnath Mazumdar\orcidlink{0000-0002-1751-2569}}
\address{Department of Digitalization, Copenhagen Business School, \\ Solbjerg Plads 3, 2000 Frederiksberg, Denmark \\
{rakr19ab@student.cbs.dk,sma.digi@cbs.dk}}

\begin{abstract}
Recent studies have revealed that when LLMs are appropriately prompted and configured, they demonstrate mixed results. Such results often meet or exceed the baseline performance. However, these comparisons have two primary issues. First, they mostly considered only reliability as a comparison metric and selected a few LLMs (such as \texttt{Codex} and \texttt{ChatGPT}) for comparision. This paper proposes a comprehensive code quality assessment framework called \textbf{P}rogrammatic \textbf{E}xcellence via \textbf{LL}M \textbf{I}teration (PELLI). PELLI is an iterative analysis-based process that upholds high-quality code changes. We extended the state-of-the-art by performing a comprehensive evaluation that generates quantitative metrics for analyzing three primary nonfunctional requirements (such as maintainability, performance, and reliability) while selecting five popular LLMs. For PELLI's applicability, we selected three application domains while following Python coding standards. Following this framework, practitioners can ensure harmonious integration between LLMs and human developers, ensuring that their potential is fully realized. PELLI can serve as a practical guide for developers aiming to leverage LLMs while adhering to recognized quality standards. This study's outcomes are crucial for advancing LLM technologies in real-world applications, providing stakeholders with a clear understanding of where these LLMs excel and where they require further refinement. Overall, based on three nonfunctional requirements, we have found that \texttt{GPT-4T} and \texttt{Gemini} performed slightly better. We also found that prompt design can influence the overall code quality. In addition, each application domain demonstrated high and low scores across various metrics, and even within the same metrics across different prompts.
\end{abstract}

\begin{keyword}
Code\sep LLMs \sep Maintainability \sep Performance \sep Reliability \sep Python
\end{keyword}
\end{frontmatter}

%%%%%%%%%%%%%%%%%%%%%%%%%%%%%%
\section{Introduction}
%%%%%%%%%%%%%%%%%%%%%%%%%%%%%%
A global software survey shows that 42\% of large IT organizations have actively deployed AI, with an additional 40\% noting they are actively exploring the technology~\cite{IBM}. In particular, specialized large language models (LLMs) for code generation are increasingly being used to optimize the productivity of software developers~\cite{poldrack2023aiassisted}. Existing research evaluating LLMs for coding reveals some prevalent trends, notably a large focus on functional correctness, which has been an important driver today's LLM research~\cite{huang2024effibench,zhong2024chatgpt,liu2023code,du2023classeval,li2022automating,chen2021evaluating,austin2021program}. However, concerns have been raised about its inability to fully represent real-world developer interactions~\cite{Liang2023can}. Therefore, some studies advocate for more complex evaluations~\cite{huang2024effibench,nascimento2023comparing,liu2023code,yang2023harnessing}. We found a gap in the current LLM-based code integration-focused research in that none of these papers comprehensively discussed a framework that \textit{can help integrate LLM-generated code in software production environment}.

\textbf{Problem Context:} Most of the scholarly works focuses on reliability~\cite{huang2024effibench,zhong2024chatgpt,poldrack2023aiassisted,mastropaolo2023robustness,bubeck2023sparks,liu2023code,du2023classeval,schäfer2023empirical,Perry_2023,li2022automating,austin2021program,pearce2021asleep,chen2021evaluating} while maintainability~\cite{poldrack2023aiassisted,mastropaolo2023robustness,bubeck2023sparks,li2022automating,chen2021evaluating} and performance efficiency~\cite{huang2024effibench,nascimento2023comparing} are overlooked despite their importance in software development. We also found that most of the earlier works primarily focused on \texttt{Codex}~\cite{mastropaolo2023robustness,Perry_2023,pearce2021asleep,chen2021evaluating} and later \texttt{ChatGPT}~\cite{zhong2024chatgpt,poldrack2023aiassisted,bubeck2023sparks,schäfer2023empirical}. Results shows, LLMs also perform poorly in real-world applications compared to controlled environments~\cite{chen2021evaluating}. These findings highlight the need for more comprehensive evaluations covering more LLMs and multiple nonfunctional requirements while focusing on more representative real-world applications to enable their effective integration into software engineering processes~\cite{wang2024software,fan2023large}.

\textbf{Research Context:} The goal of this study is to propose a framework to generate code from LLMs to accelerate overall quality software generation. For research purposes, we selected five popular LLMs~\cite{poldrack2023aiassisted,zhong2024chatgpt,chen2021evaluating,huang2024effibench,Perry_2023,pearce2021asleep}. They are \texttt{GPT-4-Turbo or GPT-4T}, \texttt{DeepSeek Coder 33B Instruct or DeepSeek Coder}, \texttt{Gemini Pro 1.0 or Gemini}, \texttt{Codex} and \texttt{CodeLLama 70b Instruct or CodeLLama}. In this paper, our focus is to answer:~\textbf{How can Large Language Models be effectively integrated into a modern software development process?} To answer this research question, we first proposed a framework called \textbf{P}rogrammatic \textbf{E}xcellence via \textbf{LL}M \textbf{I}teration or \texttt{PELLI}. \texttt{PELLI} is an iterative software development framework that provides guidelines for effectively integrating and using LLMs in practice. Second, we compare five popular LLMs based on three nonfunctional requirements, i.e., maintainability, performance, and reliability~\cite{Boehm1978} to showcase the framework applicability. Furthermore, the analysis is based on three popular application domains (such as high performance computing, machine learning, and data processing). We selected Python as our language for coding purposes while varying the prompt lengths to show code's quality variation. Through our comprehensive analysis, our contributions are:
\begin{enumerate}
    \item We proposed a framework called \textbf{P}rogrammatic \textbf{E}xcellence via \textbf{LL}M \textbf{I}teration or \texttt{PELLI} and discuss its relevance and interesting findings. 
    \item We found that LLMs with the worst memory usage matched LLMs with the best maintainability. In contrast, LLMs that demonstrate high variability in memory usage, such as \texttt{GPT-4T} and \texttt{Gemini} rank among the highest in terms of maintainability. We found that LLMs are generally well-equipped to balance code density, comment ratio, and complexity.
   \item \texttt{GPT-4T} produces reliable and efficient Python code. The convention and refactoring findings suggest that although LLMs can handle the theoretical aspects of balancing complexity, code density, and comment ratios, they struggle to apply these principles within specific programing languages due to language-dependent nuances.
   \item Prompt length can affect error rates; however, the characteristics of individual LLMs also play critical roles in determining the reliability of the generated code.
   \item Longer prompt solutions perform slightly below medium prompt solutions, suggesting more dense complexity and less dispersion. As prompts become more detailed, the variability in potential solutions also increases, reflecting the LLMs' adaptive responses to the expanding requirements.
\end{enumerate}
%
%%%%%%%%%%%%%%%%%%%%%%%%%%%%%%%%%%
%%%%%%%%%%%%%%%%%%%%%%%%%%%%%%%%%%
\section{Related Work}
\label{sec:related_work}
%%%%%%%%%%%%%%%%%%%%%%%%%%%%%%%%%%
%%%%%%%%%%%%%%%%%%%%%%%%%%%%%%%%%%
Selected studies based on the generated LLM codes are categorized based on three nonfunctional requirements, i.e., maintainability, reliability, and performance.
%%%%%%%%%%%%%%%%%%%%%%%%%%%%%%%%%%
\subsection{Reliability}
%%%%%%%%%%%%%%%%%%%%%%%%%%%%%%%%%%
Reliability is the most popular nonfunctional requirement in LLM research~\cite{poldrack2023aiassisted,expectation_vs_experience,bubeck2023sparks,mastropaolo2023robustness}. In these studies, a common focus was the assessment of functional correctness using benchmark datasets. HumanEval~\cite{chen2021evaluating} and Basic Programming Problems (MBPP)~\cite{austin2021program} are popular benchmarks for evaluating language comprehension, algorithms, and basic mathematics to determine functional correctness in generated code. However, these benchmarks include insufficient tests and imprecise problem descriptions~\cite{liu2023code}. Chen et al. revealed limitations in \texttt{Codex}, especially in handling complex operation chains, which indicated limitations in terms of reliability for more involved tasks~\cite{chen2021evaluating}. Many benchmarking datasets typically focus on function-specific problem descriptions, issue~\cite{du2023classeval} addresses with the introduction of ClassEval. The benchmark on class-level code generation reveals that LLMs perform significantly worse when generating class implementations than standalone methods. These findings highlight the weaknesses of functionality-specific benchmarks identified in previous research~\cite{chen2021evaluating,austin2021program,liu2023code} and question the reliability of LLM-generated code. Li et al. also proposed an alternative benchmark to evaluate key code review tasks~\cite{li2022automating}. Like~\cite{huang2024effibench} benchmark for performance efficiency,~\cite{zhong2024chatgpt} attempts to assist with the evaluation of LLM reliability using a benchmark comprising 1208 StackOverflow coding questions. By applying their benchmark on various LLMs,~\cite{zhong2024chatgpt} finds various patterns in the generated code that lead to unreliability and security concerns. Consequently, Zhong et al.~\cite{zhong2024chatgpt} is the only study explicitly addressing reliability. As demonstrated, most reliability research is rooted in the efficient evaluation of LLMs using functional correctness, which is a useful tool for evaluating and improving LLMs for code, essentially making them more reliable. Schäfer et al. analyzed reliability from a more applied perspective and demonstrated that LLMs can be leveraged to generate more reliable solutions~\cite{schäfer2023empirical}. They performed this using an adaptive \texttt{GPT-3.5T}-based test generation tool that outperformed all existing state-of-the-art methods. Only one of the included reliability references explicitly attempted to evaluate this. However, its contribution is a benchmark dataset rather than a thorough empirical analysis. This highlights the importance of addressing reliability more directly, particularly in practical contexts.

%%%%%%%%%%%%%%%%%%%%%%%%%%%%%%%%%%
\subsection{Maintainability}
%%%%%%%%%%%%%%%%%%%%%%%%%%%%%%%%%%
Poldrack et al. report that \texttt{GPT-4T} effectively simplifies and standardizes the source code, enhancing its maintainability through simulations of real-world developer interactions~\cite{poldrack2023aiassisted}. A related study found that programmers often struggled to understand and edit generated code~\cite{expectation_vs_experience}, which is in contrast to the findings of~\cite{poldrack2023aiassisted} and underscores concerns about maintainability. Bubeck et al.~\cite{bubeck2023sparks} observes that generated code by~\texttt{GPT-4T} may be challenging to maintain due to frequent syntactic errors and semantic flaws. Mastropaolo et al. investigated the impact of prompting on code generation with \texttt{Codex} and found that varying descriptions of natural language result in 46\% different outputs despite prompting including identical functionality requests~\cite{mastropaolo2023robustness}. Although the aforementioned studies indicate both maintainable~\cite{poldrack2023aiassisted} and less maintainable code~\cite{expectation_vs_experience,bubeck2023sparks,mastropaolo2023robustness} for LLM-generated code, none are explicitly designed with a focus on systematically evaluating the overall maintainability.

%%%%%%%%%%%%%%%%%%%%%%%%%%%%%%%%%%
\subsection{Performance}
%%%%%%%%%%%%%%%%%%%%%%%%%%%%%%%%%%
Performance efficiency is one of the most neglected categories in LLM research. Nascimento et al. found that while \texttt{ChatGPT} can outperform novice programmers, it struggles against more experienced ones~\cite{nascimento2023comparing}. Similarly,~\cite{Liang2023can} found that ChatGPT-4 produces correct code for high-level structure, but can contain errors in its lower-level implementation. Next, the analysis by Huang et al. reveals that \texttt{GPT-4T} produces the most efficient code~\cite{huang2024effibench}. Performance efficiency-focused studies are sparse and often rely on similar methodologies, as seen in~\cite{huang2024effibench,nascimento2023comparing}. A notable threat to the validity of these studies is the potential inclusion of LeetCode\footnote{https://leetcode.com/problemset/} problems in the LLM training datasets. The specific characteristics of these problems are useful for the instruction training of LLMs, and considering how readily available these problems are, it is not unreasonable to expect. This highlights the critical need for an analysis that assesses the performance efficiency of alternative code bases.

%%%%%%%%%%%%%%%%%%%%%%%%%%%%%%%%%%
\section{Proposed Framework: PELLI}
\label{sec:PELLI}
%%%%%%%%%%%%%%%%%%%%%%%%%%%%%%%%%%
%
\begin{figure}[!hbt]
    \centering
    \includegraphics[width=\linewidth]{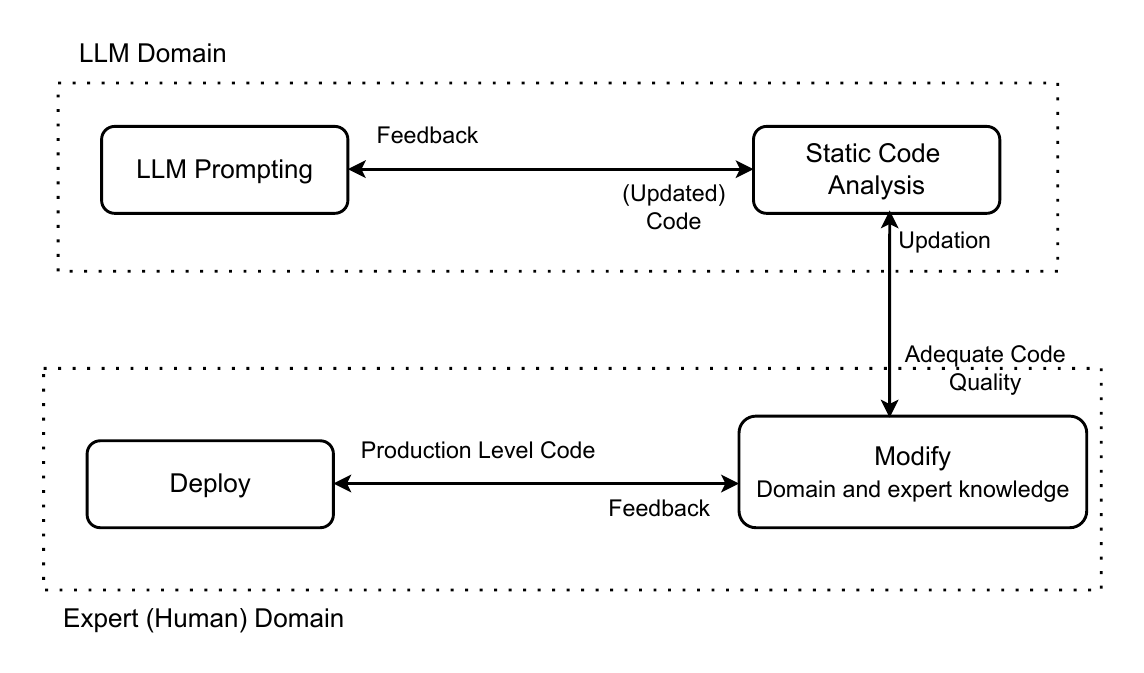}
    \caption{Representation of proposed PELLI framework}
    \label{fig:pelli}
\end{figure}

To assist practitioners in their endeavors to adopt LLMs in their software development processes and contribute to the open problem of effective integration identified by~\cite{fan2023large}, this research introduces the \textbf{P}rogrammatic \textbf{E}xcellence via \textbf{LL}M \textbf{I}teration or \texttt{PELLI}. The key components of \texttt{PELLI} include iterative processes and systematic code evaluation. First, iterative refinement of software is a well-established practice~\cite{AlSaqqa2020AgileSD,continuous_software_engineering,iterative_enhancement}, with \texttt{PELLI} incorporating similar iterative methods to ensure continuous improvement and adaptation of the software development process. Second, the importance of systematic evaluation cannot be overstated. Previous research has demonstrated that LLMs may generate erroneous code~\cite{Liang2023can,pearce2021asleep}, a problem that tends to prevail when humans interact with such systems. This issue often arises from users overly trusting the output without sufficient critical evaluation~\cite{Perry_2023,chen2021evaluating}. To mitigate these risks, \texttt{PELLI} emphasizes the need for rigorous and methodical code evaluation to ensure maintainability, performance, and reliability. Figure~\ref{fig:pelli} presents the framework. It has two primary components i.e., LLM-Domain and Expert-Domain which are discussed next.
%%%%%%%%%%%%%%%%%%%%%%%%%%%%%%%%%%
\subsection{LLM Domain}
%%%%%%%%%%%%%%%%%%%%%%%%%%%%%%%%%%

%%%%%%%%%%%%%%%%%%%%%%%%%%%%%%%%%%
\subsubsection{LLM Prompting}
%%%%%%%%%%%%%%%%%%%%%%%%%%%%%%%%%%
%
\begin{figure*}[ht]
\centering
\includegraphics[width=\textwidth]{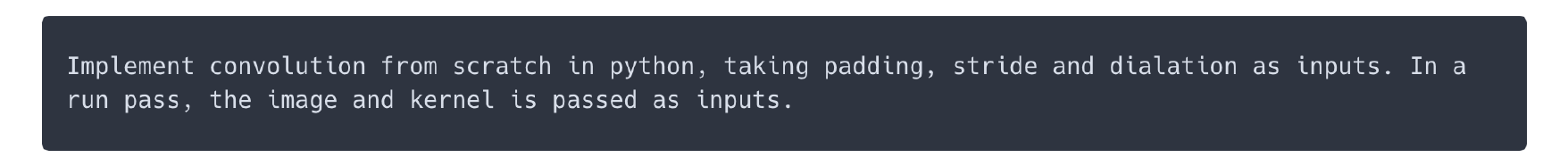}
\caption{Example of short prompt for Convolutional neural network code generation}
\label{fig:short-prompt-example}
\end{figure*}
\begin{figure*}[ht]
    \centering
    \includegraphics[width=\textwidth]{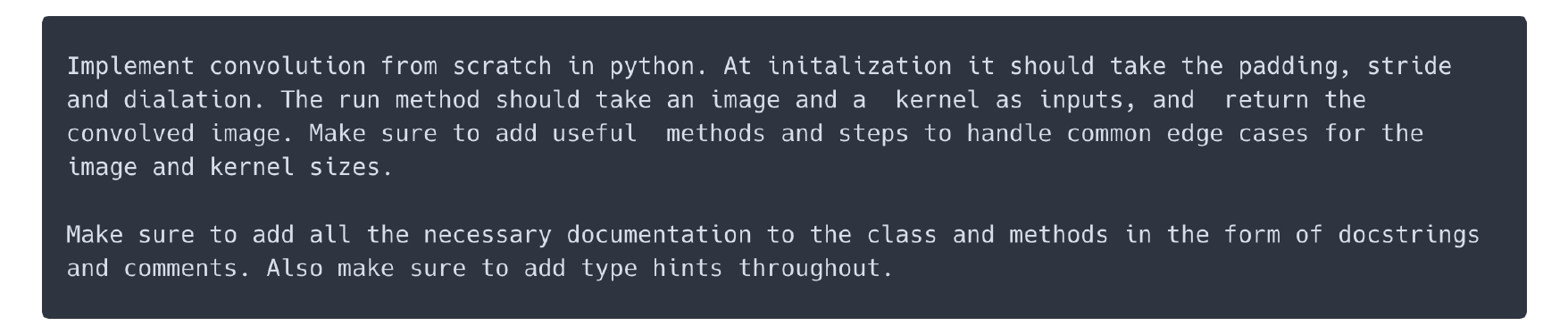}
    \caption{Example medium prompt for Convolutional neural network code generation}
    \label{fig:medim-prompt-example}
\end{figure*}
\begin{figure*}[ht]
    \centering
    \includegraphics[width=\textwidth]{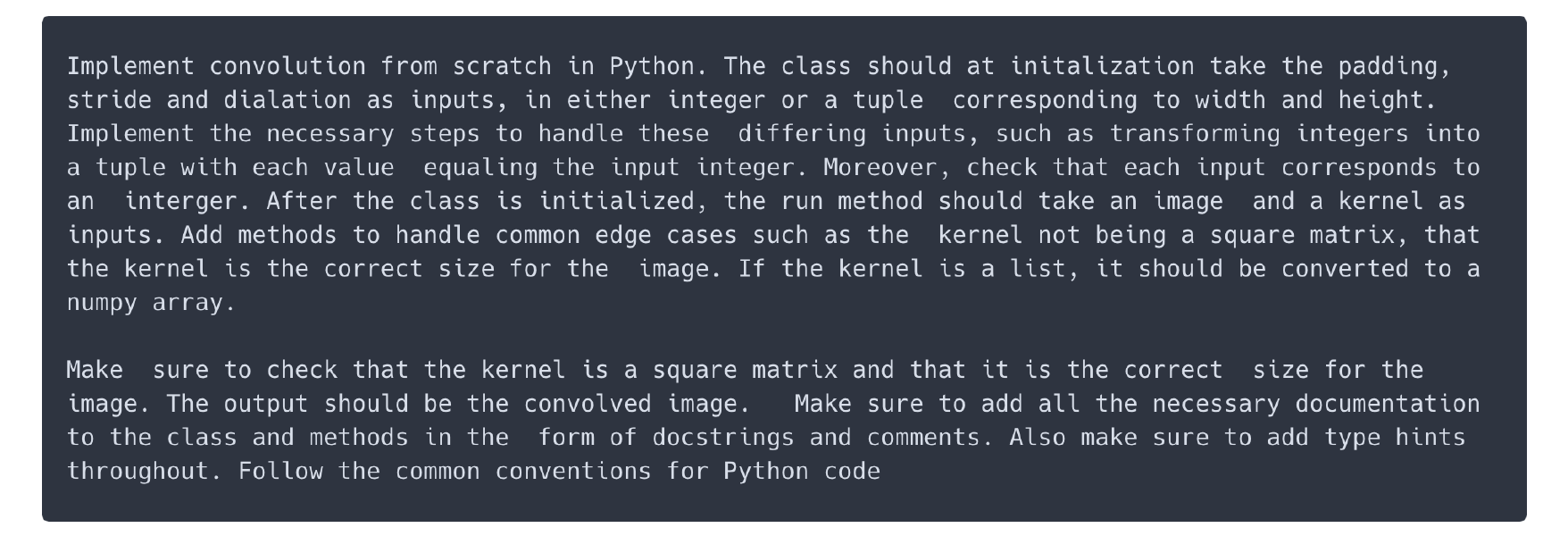}
    \caption{Example long prompt for Convolutional neural network code generation}
    \label{fig:long-prompt-example}
\end{figure*}
The \textbf{LLM Prompting} component is the first component that forms the foundation of the \texttt{PELLI} framework. A prompting methodology is employed to reflect the various interactions that will similarly occur in practice. This inclusion is strongly motivated by the findings~\cite{mastropaolo2023robustness}, with the difference here being the addition of a more defined methodology to ensure generalizability. Here, developers are tasked with describing their requirements using a useful prompt that is subsequently passed to the LLM for inference. As found throughout the analysis, the prompt's quality greatly influences the generated code. While LLMs may improve over time, becoming better at adhering and generating more thorough code using fewer details, the analysis found that detailed prompts that are neither overly sparse nor excessively verbose are the best at their current capacity. More generally, developers must be considerate of their descriptions and specifications in their prompts, which is further emphasized by the vast discrepancy between outputs for different prompts that otherwise requested the same functionality found by~\cite{expectation_vs_experience}. A stratified prompting approach was used in this research and classified by prompt length: short, medium, and long, based on the findings by~\cite{mastropaolo2023robustness}.

\textbf{Short Prompts} are characterized by requesting only the implementation of an algorithm from scratch with a minimal context, relying on LLMs to infer essential requirements and adhere to best practices. This approach is critical for testing the ability of LLMs to abstract and apply coding principles outside the prompts. A concern related to this prompt is that the minimal context can introduce ambiguity, potentially resulting in code that does not satisfy the intended algorithmic functionality.

\textbf{Medium Prompts} specify functionality, use cases, documentation, and type hints; however, they lack detailed guidance on argument handling and data types. The expectation is that LLMs engage in creative problem-solving and interpret instructions to produce code that aligns with specified details. Ideally, this should foster the development of more robust and adaptable solutions, as the LLMs must infer the best practices for data handling and argument processing based on the context provided.

\textbf{Long Prompt} includes all the requirements for medium prompts and additional instructions on handling edge cases, unspecified arguments, and specific data types. They are designed to minimize ambiguity and direct the LLM toward producing the best code solutions, thus eliminating any reliance on the LLM's reasoning abilities. This approach tests the capacity of LLM to process complex instructions and craft solutions that meet specific criteria.

The stratification of prompt lengths serves two main research goals. First, it evaluates LLMs' adaptability to varying levels of detail in the instructions, which is crucial to cover diverse scenarios. Second, the proposed model provides insights into LLM biases, coding conventions, and the ability to go beyond prompt constraints to achieve higher quality. Additionally, this method recognizes the iterative nature of coding, where initial drafts are refined over time, and evaluates the LLM's effectiveness throughout this process. If an LLM did not produce an acceptable response, follow-up prompts without additional outcome information. All LLMs were also prompted without prior context to ensure that the details from earlier prompts did not influence their responses. It should be noted that using prompts of varying lengths and detail levels can lead to diverse usability and functionality outcomes. Longer and more detailed prompts are expected to yield solutions with greater functionality; however, this has not been explicitly tested. Ideally, each LLM should produce a comprehensive solution that meets general code quality standards, regardless of the prompt's detail level. The variation in the prompt detail also mirrors different user needs and intentions, essentially exploring different developers’ preferences with the expectation that all solutions will maintain high quality irrespective of the prompt's comprehensiveness.

%%%%%%%%%%%%%%%%%%%%%%%%%%%%%%%%%%
\subsubsection{Static Code Analysis}
%%%%%%%%%%%%%%%%%%%%%%%%%%%%%%%%%%
The following is the \textbf{Analysis} phase, where the synthesized LLM code is thoroughly evaluated for quality and relevance. This component is pivotal because it affects subsequent steps in the development process. This analysis evaluates the code against various metrics, particularly Python language-specific standards. In this area, LLMs showed increased variability. For this study, we used three open source tools such as Pylint (version~2.16.2)~\cite{Gulabovska2019SurveyOS}, Radon (version~5.1.0)\footnote{\url{https://radon.readthedocs.io/en/latest/}}, and Python system and process utilities (or psutil version~5.9.0)\footnote{\url{https://psutil.readthedocs.io/en/latest/}}.

%%%%%%%%%%%%%%%%%%%%%%%%%%%%%
\textbf{Pylint}
%%%%%%%%%%%%%%%%%%%%%%%%%%%%%
Pylint is a static Python code analyzer. It checks programmatic and stylistic errors, enforces coding standards, and detects poor quality code~\cite{Pylint}. It classifies issues into five categories. They are \textit{i) Style Convention errors (C)}: Coding style violations per \textit{PEP~8}; \textit{ii) Refactoring checks (R)}: Code refactoring suggestions; \textit{iii) Warnings (W)}: Potentially problematic code snippets; \textit{iv) Errors (E)}: Code likely to cause runtime issues; and \textit{v) Fatal errors (F)}: Severe issues halting analysis. During our analysis, two style messages, \texttt{line-too-long (C0301)} and \texttt{trailing-whitespace (C0303)}, were disabled to avoid UI-related formatting discrepancies.

%%%%%%%%%%%%%%%%%%%%%%%%%%%%%
\textbf{Radon}
%%%%%%%%%%%%%%%%%%%%%%%%%%%%%
Radon computes four software metrics. They are \textit{i) Cyclomatic Complexity}: Measures decision points in the code~\cite{1702388}; \textit{ii) Maintainability Index}: Considers complexity, size, and documentation~\cite{welker2001software}; \textit{iii) Number of Delivered Bugs}: Predicts potential reliability issues~\cite{halstead1977elements}; and \textit{iv) Source code statistics}: Includes lines of code (LOC) and comments-to-code ratio.

%%%%%%%%%%%%%%%%%%%%%%%%%%%%%
\textbf{psutil}
%%%%%%%%%%%%%%%%%%%%%%%%%%%%%
A monitoring tool that provides detailed information about resource utilization (such as CPU, memory, disk activity, network interfaces, and more). In this study, the CPU and memory use data were specifically collected. Monitoring with psutil requires the execution of solutions to observe system performance in real time.

As mentioned, the iterations were at the core of \texttt{PELLI}, which is prevalent among all components. Here, the \textbf{Feedback} that follows the analysis uses learnings from the analysis and iteratively includes these learnings in refinement prompts intended to augment the non-expertise-related aspects of the code. Although not explicitly applied in this paper, previous research has found that LLM output can be improved and often requires human feedback through iterative refinement~\cite{poldrack2023aiassisted,nascimento2023comparing,austin2021program}.

Finally, any critical errors are addressed with follow-up prompts and adjustments to ensure that the LLM-generated code is executable. These minor adjustments include adding library imports, assigning variables to class attributes, and transposing matrices. In the event of major changes, LLMs must generate alternative solutions. This process does not confirm that the implemented algorithms worked as intended; thus, while they are capable of execution, they may be erroneous.

%%%%%%%%%%%%%%%%%%%%%%%%%%%%%%%%%%
\subsection{Expert Domain}
%%%%%%%%%%%%%%%%%%%%%%%%%%%%%%%%%%

%%%%%%%%%%%%%%%%%%%%%%%%%%%%%%%%%%
\subsubsection{Modify}
%%%%%%%%%%%%%%%%%%%%%%%%%%%%%%%%%%
When the code is deemed adequate (refer to Figure~\ref{fig:pelli}), the process moves to the \textbf{Modify} component. \textbf{Modify} component is where human developers come into play, leveraging domain-specific knowledge to refine and optimize the LLM-generated code. LLMs demonstrate variability in performance across different domains, and previous studies have found that certain LLMs are inadequate to handle more complex scenarios~\cite{bubeck2023sparks,poldrack2023aiassisted}, making this step is crucial for ensuring a sufficient level of quality. The results demonstrate that the human benchmark was exceeded, particularly in terms of performance efficiency. However, domain differences were also observed. By having developers attend to these areas of the source code, best practices can be improved, leading to higher output quality. One concern is found by~\cite{expectation_vs_experience}, where developers occasionally found the code generated by LLMs difficult to understand, debug, and generally edit. If the components within the nonexpert domain are performed adequately, many of these concerns should be minimized using the added documentation and generally descent code style. However, some are inevitable. Both~\cite{poldrack2023aiassisted} and~\cite{bubeck2023sparks} found that LLMs are also useful for understanding code. Although not included directly in \texttt{PELLI}, developers are encouraged to exploit such strengths to improve their understanding of the code.

Naturally, any changes should be similarly analyzed using the same methods applied for the LLM-generated code, enabling comparison and further understanding of whether the desired effects were achieved. This process is similar to the feedback loop in the \textit{LLM Domain}. It is highly iterative, and developers are expected to iterate over their improvement, leveraging LLMs for useful new code generation and expertise domains to attend to common weaker areas.

%%%%%%%%%%%%%%%%%%%%%%%%%%%%%%%%%%
\subsubsection{Deploy}
%%%%%%%%%%%%%%%%%%%%%%%%%%%%%%%%%%
Finally,~\textbf{Deploy} phase involves deploying/executing the code to assess its real-world functionality and performance. Common for most software development processes is the inclusion and execution of unit tests, which LLMs have similarly proved proficient at \cite{schäfer2023empirical}. 

\texttt{PELLI} emphasizes the importance of thorough testing to ensure that collaborative efforts between LLMs and humans exceed those of non-LLM assisted human developers. Therefore, any detected discrepancies in the unit tests are addressed through domain-specific modifications, which are illustrated by the feedback relation between \textbf{Deploy} and \textbf{Modify}. When the source code is finally deemed adequate, it can be deployed in a confident manner.

%%%%%%%%%%%%%%%%%%%%%%%%%%%%%%%%%%%%%%%%%%%%%%%%%%%%%%%%%%%%%%%%%%%%
\section{Methodology of Analysis}
%%%%%%%%%%%%%%%%%%%%%%%%%%%%%%%%%%%%%%%%%%%%%%%%%%%%%%%%%%%%%%%%%%%%

%%%%%%%%%%%%%%%%%%%%%%%%%%%%%%%%%%
\subsection{Use case} 
\label{sec:application_domain_selection}
%%%%%%%%%%%%%%%%%%%%%%%%%%%%%%%%%%
Nine representative algorithms across three distinct application domains (such as high-performance computing (HPC), machine learning (ML), and data processing) are chosen to reflect the variation of real-world applications. \textit{HPC} tasks focused algorithms are \textit{quick sort (QS)}, \textit{Strassen matrix multiplication (SMM)} and \textit{Monte Carlo simulation (MCS)}. MCS is used for financial modeling and challenges LLMs in probabilistic scenarios and large iterations. 

\textit{ML} covers many applications, ranging from predictive analytics to autonomous systems. The selected ML algorithms are the following: \textit{attention}, the arguably most influential algorithm in the recent times~\cite{NIPS2017_3f5ee243}. This model is considered critical in natural language processing research. Without this algorithm, conversational LLMs would not have been feasible. For the \textit{Attention} mechanism to function effectively, it requires the management of various inter-dependencies, which requires LLMs to track and execute these efficiently. \textit{Convolution} algorithm is essential for image processing and computer vision. It involves explicit spatial and mathematical operations. \textit{Principal component analysis (PCA)} is a dimensional reduction algorithm that exposes the information that underlines the datasets and explains most of the variability with a smaller number of variables. 

\textit{Data processing} tasks are essential for making raw data informative. The selected algorithms include: \textit{Huffmann coding (HC)}, a widely used compression algorithm;~\textit{PageRank (PR)} is an algorithm that processes data in a graph-matrix structure and~\textit{Rabin-Karp (RK)} is a string pattern matching algorithm useful in many pattern matching applications.

%%%%%%%%%%%%%%%%%%%%%%%%%%%%%%%%%%
\subsection{Selected Metrics}
%%%%%%%%%%%%%%%%%%%%%%%%%%%%%%%%%%
%
\begin{table}[!hbt]
    \centering
    \caption{Eleven metrics collected from three open source tools for analyzing three primary nonfunctional requirements}
\label{tab:iso_5055_metrics}
\begin{adjustbox}{width=1.0\columnwidth}
\begin{tabular}{|c|c|c|}
\hline
\textbf{Nonfunctional Requirements}        & \textbf{Metric} & \textbf{Tool} \\ \hline 
        \multirow{5}{*}{Maintability}           & Maintainability Index (MI) & Radon  \\
                                                & Convention (C)            & Pylint \\
                                                & Refactor (R)              & Pylint \\
                                                & SLOC-to-Methods (SLOC-to-M)      & Radon  \\
                                                & Comments-to-LOC (C-to-LOC)       & Radon  \\\cline{1-3} 
        \multirow{2}{*}{Performance} & CPU Usage (CPU)             & psutil \\
                                                & Memory Usage (Memory)          & psutil \\\cline{1-3} 
        \multirow{4}{*}{Reliability}            & Cyclomatic Complexity (CC) & Radon  \\
                                                & Delivered Bugs (Bugs)        & Radon  \\ 
                                                & Warnings              & Pylint \\
                                                & Errors                & Pylint \\\cline{1-3} 
    \end{tabular}
\end{adjustbox}
\end{table}
Metrics from the open-source tools are organized to analyze three primary nonfunctional requirements (such as maintainability, performance, and reliability, see Table~\ref{tab:iso_5055_metrics}). Python-specific maintability metrics included Pylint's style and refactoring messages, the source lines of code (SLOC) for each method and the comments-to-code ratio. In this study, the CPU and memory usage was specifically collected from psutil. Reliability was assessed according to cyclomatic complexity and the number of delivered bugs. Pylint and Radon contributed to the code reliability evaluation by checking for warnings and errors.

%%%%%%%%%%%%%%%%%%%%%%%%%%%%%%%%%%%%%%%%%%%%%%%%%%%%%%%%%%%%%%%%%%%%
\subsection{Data Cleaning and Preprocessing}
\label{sec:Data Cleaning and Preprocessing}
%%%%%%%%%%%%%%%%%%%%%%%%%%%%%%%%%%%%%%%%%%%%%%%%%%%%%%%%%%%%%%%%%%%%
%
\begin{figure}[ht]
    \centering
    \includegraphics[width=\linewidth]{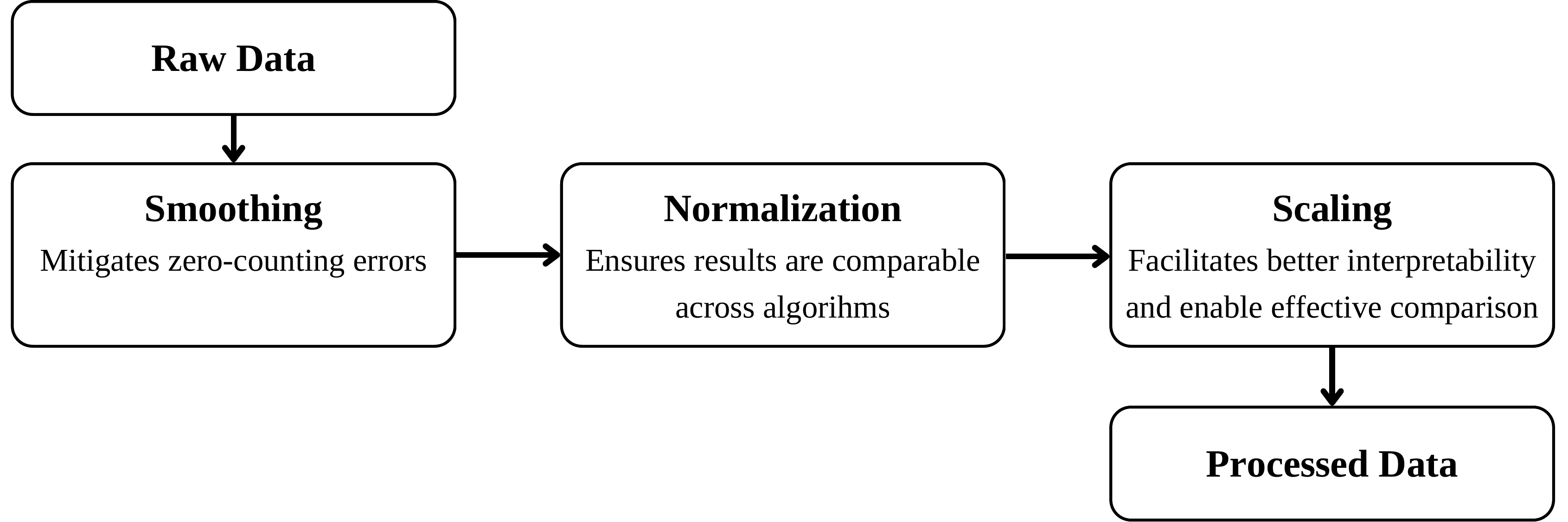}
    \caption{Representation of data cleaning and preprocessing stages}
    \label{fig:processing_process}
\end{figure}
Variability in prompt length and detail can lead to differences in code output and potential style violations, thereby complicating direct comparisons. Longer prompts can result in more coding errors, skewing analysis, and biased assessments. To address this issue, preprocessing can normalize the results across different code lengths, which ensures fair comparisons. The motivation for the preprocessing step was to provide a clear, context-relative understanding of the data that, when later merged with algorithms in the same domain, was not biased toward the produced code. Table~\ref{tab:metric_preprocessing} summarizes the pre-processing for each metric.

\textbf{Smoothing}
To mitigate zero-counting errors, we apply $add$-$k$ smoothing, which is a common modification used to mitigate zero-counting errors~\cite{Jurafsky_Martin_2024}. An appropriate value for $k$ can be determined using various methods; however, we set $k$ to $0.01$. It does not apply unjust violations to the solutions and stabilizes the final evaluation.

\textbf{Normalization}
Normalization is applied (refer to Equation~\ref{eq:normalization}) for count-based code violations by dividing each raw count by the total lines of code in the solution, which reflects violations per line of code, ensuring comparable results across various prompt lengths.
\begin{equation}
\label{eq:normalization}
    \text{Normalized Violations} = \frac{\text{Number of Violations}}{\text{Total Lines of Code}}
\end{equation}
The SLOC and Cyclomatic complexity (CC) can be normalized using many methods. The proposed method assesses how well logic and structural complexity are dispersed across methods rather than just lines of code.

\textbf{Scaling}
Scaling standardizes metrics to improve interpretability and comparison (see Equation~\ref{eq:scaled_equation}, where {$x_{a}$ is a vector of all observed values for a metric related to algorithm $a$}). Each metric is scaled to a range between zero and one, transforming absolute measurements into relative values.
\begin{equation}
\label{eq:scaled_equation}
\text{Scaled Values} = \frac{x_{a}}{\text{max}(x_{a})}
\end{equation}
For fixed-range metrics like the maintainability index ($0-100$), the upper range is used for scaling. Metrics where higher values indicate poorer performance are inverted by subtracting each from one and aligning all metrics such that values closer to one represent better performance.
\begin{table}[ht]
\centering
\caption{Summarizing each metrics of pre-processing.(MO: means Max Observed)}
\label{tab:metric_preprocessing}
\begin{adjustbox}{width=0.85\columnwidth}
\begin{tabular}{|c|c|c|c|c|}
    \hline
     \textbf{Metric} & \textbf{Smoothing} & \textbf{Norm.} & \textbf{Scaling} & \textbf{Inverse}   \\
    \hline
    \begin{tabular}[c]{@{}c@{}}Maintainability\\Index\end{tabular}    & -    & -          & 10            & False  \\ \hline
    Convention              & 0.01 & LOC        & MO  & True \\ \hline
    Refactoring             & 0.01 & LOC        & MO  & True \\ \hline
    Comments                &   -  & LOC        & MO  & False \\ \hline
    SLOC                    &   -  & Methods    & MO  & True \\ \hline
    CPU Usage               &   -  & -          & 100           & True \\ \hline
    Memory Usage            &   -  & -          & MO  & True \\ \hline
    \begin{tabular}[c]{@{}c@{}}Cyclomatic\\Complexity\end{tabular}   &   -  & Methods    & MO  & True \\ \hline
    Delivered Bugs          &  -   & LOC        & MO  & True \\ \hline
    Warnings                & 0.01 & LOC        & MO  & True \\ \hline
    Errors                  & 0.01 & LOC        & MO  & True \\ \hline
\end{tabular}
\end{adjustbox}
\end{table}
%       
%%%%%%%%%%%%%%%%%%%%%%%%%%%%%%%%%%%%%%%%%%%%%%%%%%%%%%%%%%%%%%%%%%%%
\section{Results Analysis}
%%%%%%%%%%%%%%%%%%%%%%%%%%%%%%%%%%%%%%%%%%%%%%%%%%%%%%%%%%%%%%%%%%%%
%%%%%%%%%%%%%%%%%%%%%%%%%%%%%%%
\subsection{Environment Setup}
%%%%%%%%%%%%%%%%%%%%%%%%%%%%%%%
An Apple Macbook Pro with an M2 Pro chip and 16 GB of RAM running MacOS (Sonoma version 14.5) was used for the evaluation. The computer was equipped with Python v3.11, Numpy v1.24.4, SciPy v1.11.1, and Pandas v1.5.3 for evaluation. The baseline solutions were generated using the Visual Studio Code editor. Due to the considerable computational requirements of these LLMs, a virtual machine equipped with an Nvidia A40 graphics processing unit with 40 GB of RAM was also used. The LLMs were asked to implement solutions from scratch. Given that every LLM is prompted to implement each algorithm three times for the different prompt lengths. To mitigate the influence of external factors on the measured metrics, a looping process was performed in which each solution was executed five times.

%%%%%%%%%%%%%%%%%%%%%%%%%%%%%%%
\subsection{Maintainability}
\label{sec:results_maintainability}
%%%%%%%%%%%%%%%%%%%%%%%%%%%%%%%
Here, the maintainability index is first assessed, followed by the Python-specific maintainability metrics (based on PEP 8 style conventions and refactorings), and the comments-to-code ratio and SLOC for per method are evaluated.

%%%%%%%%%%%%%%%%%%%%%%%%%%%%%%%
\subsubsection{Maintainability Index}
\label{sec:maintainability_index}
\begin{figure}[ht]
    \centering
    \includegraphics[width=\columnwidth]{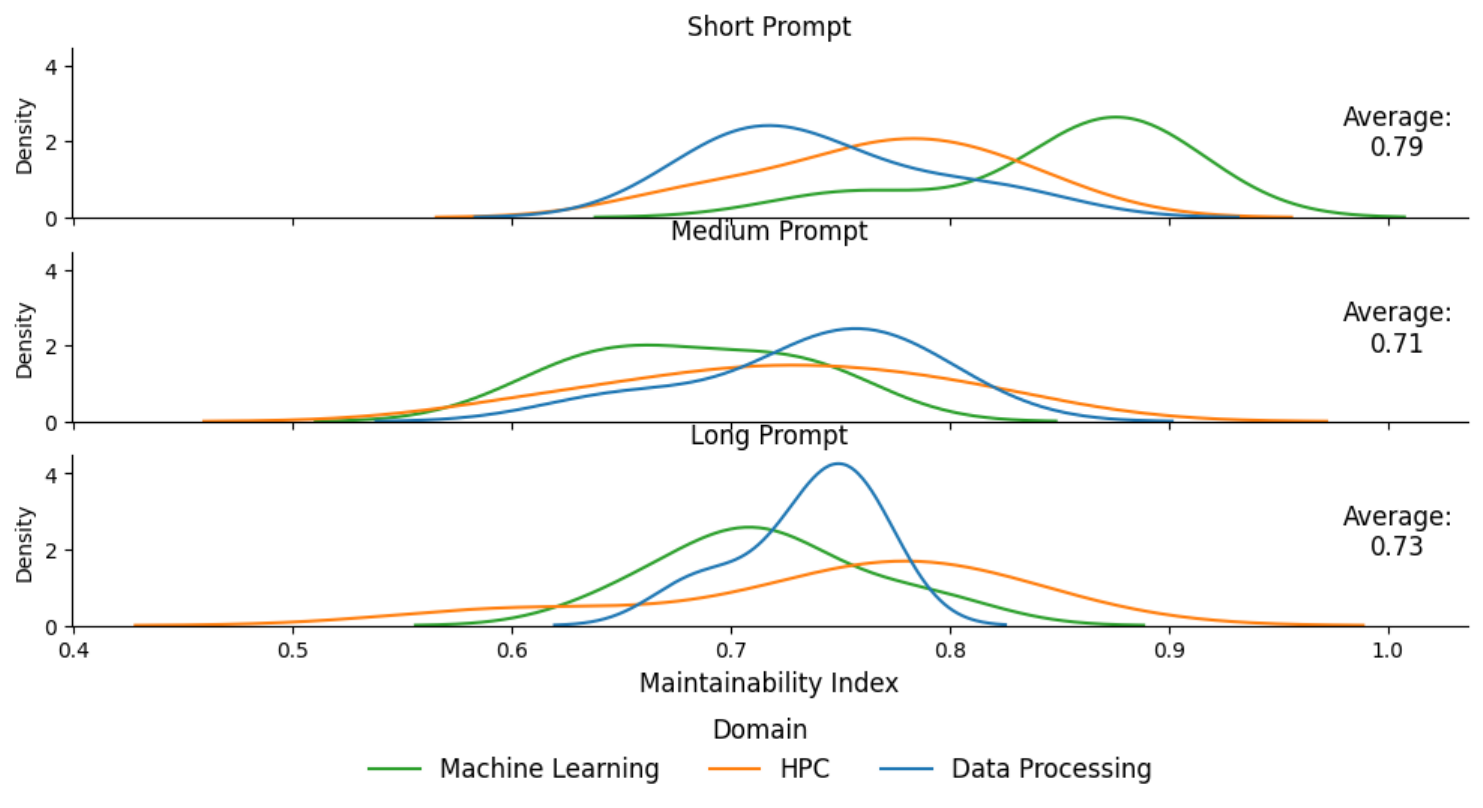}
    \caption{Maintainability Index with varying prompts and use cases}
    \label{fig:mi_domain}
\end{figure}

\begin{figure}[ht]
    \centering
    \includegraphics[width=1\columnwidth]{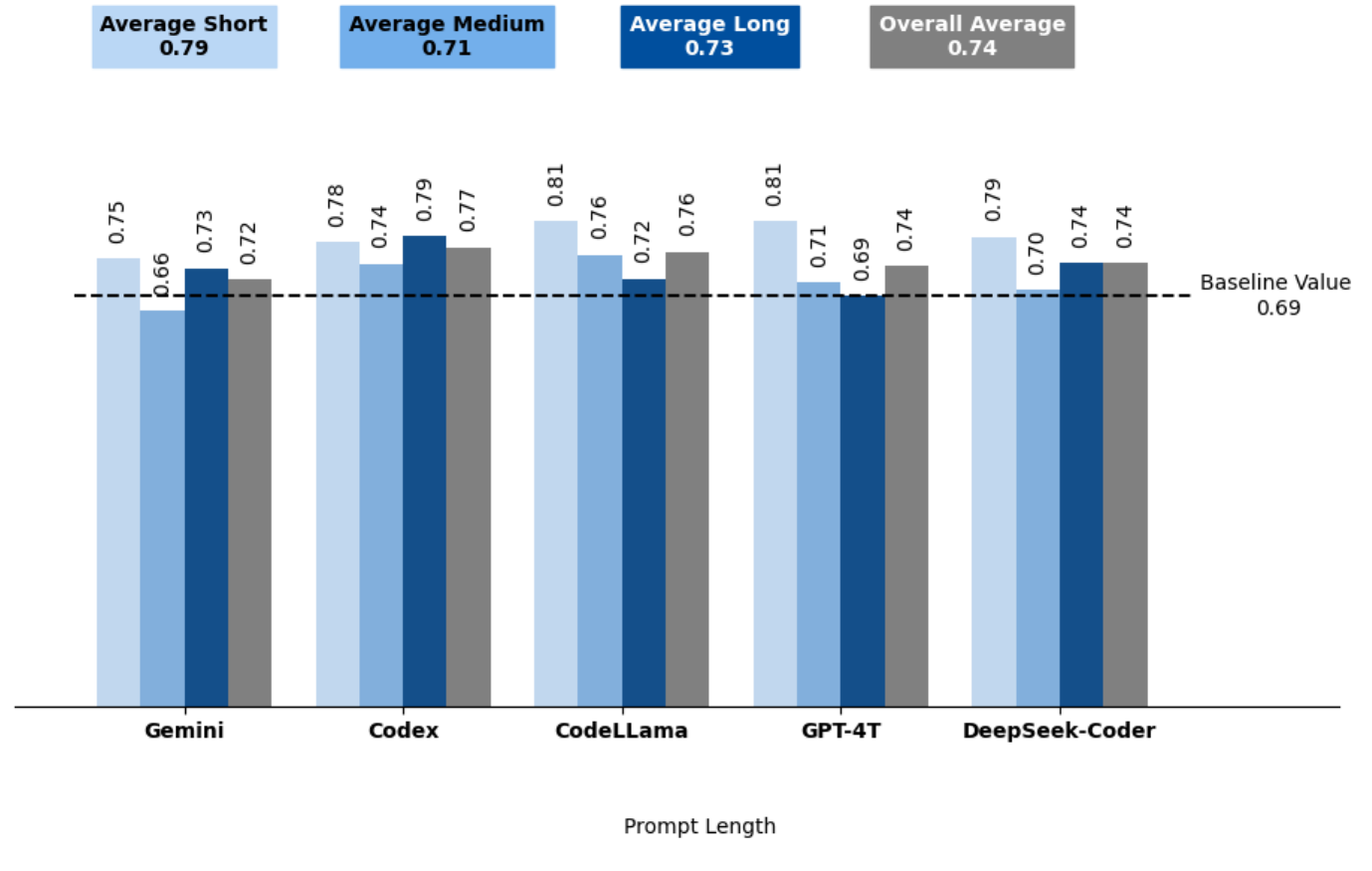}
    \caption{Measuring LLMs against Maintainability Index}
    \label{fig:mi_llm}
\end{figure}

%%%%%%%%%%%%%%%%%%%%%%%%%%%%%%%
The distribution of the maintainability index across various domains (refer to Figure~\ref{fig:mi_domain}) shows that LLMs generally exhibit consistent performance, with similar averages across prompts a relatively uniform dispersion in each domain. However, a significant outlier was observed in the ML domain, where short prompt solutions outperformed longer ones. This suggests that the brevity of these prompts may facilitate simpler solutions characterized by lower code density and reduced complexity. However, this pattern does not extend to other domains, where performance remains consistent across various prompts, which indicates possible ML domain-specific variation.
Consistency also extended to individual LLMs (see Figure~\ref{fig:mi_llm}), where LLMs tend to achieve higher scores for short prompts, which aligns with earlier observations. Although no LLM significantly underperforms, \texttt{Gemini} records the lowest scores, particularly for medium prompt solutions. All LLMs outperformed the baseline models, which suggests that the LLMs are generally well-equipped to balance code density, comment ratio, and complexity.
%

%%%%%%%%%%%%%%%%%%%%%%%%%%%%%%%
\subsubsection{Python Coding Convention Violations}
\label{sec:Python Coding Violations}
%%%%%%%%%%%%%%%%%%%%%%%%%%%%%%%
%
\begin{figure}[!hbt]
    \centering
    \includegraphics[width=\columnwidth]{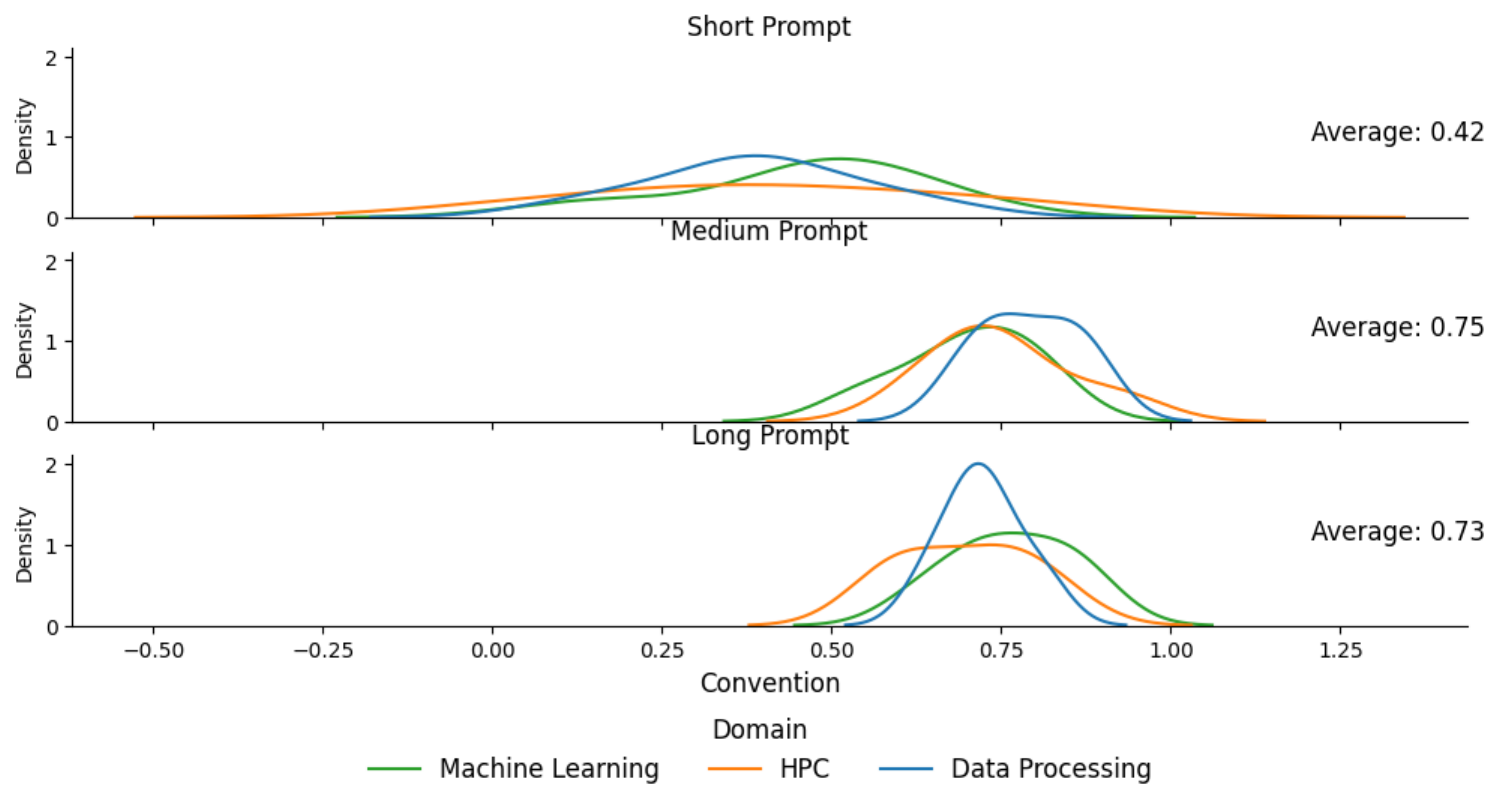}
    \caption{Style convention violations with varying prompts and use case}
    \label{fig:code_style_violations}
\end{figure}
\begin{figure}[!hbt]
    \centering
    \includegraphics[width=\columnwidth]{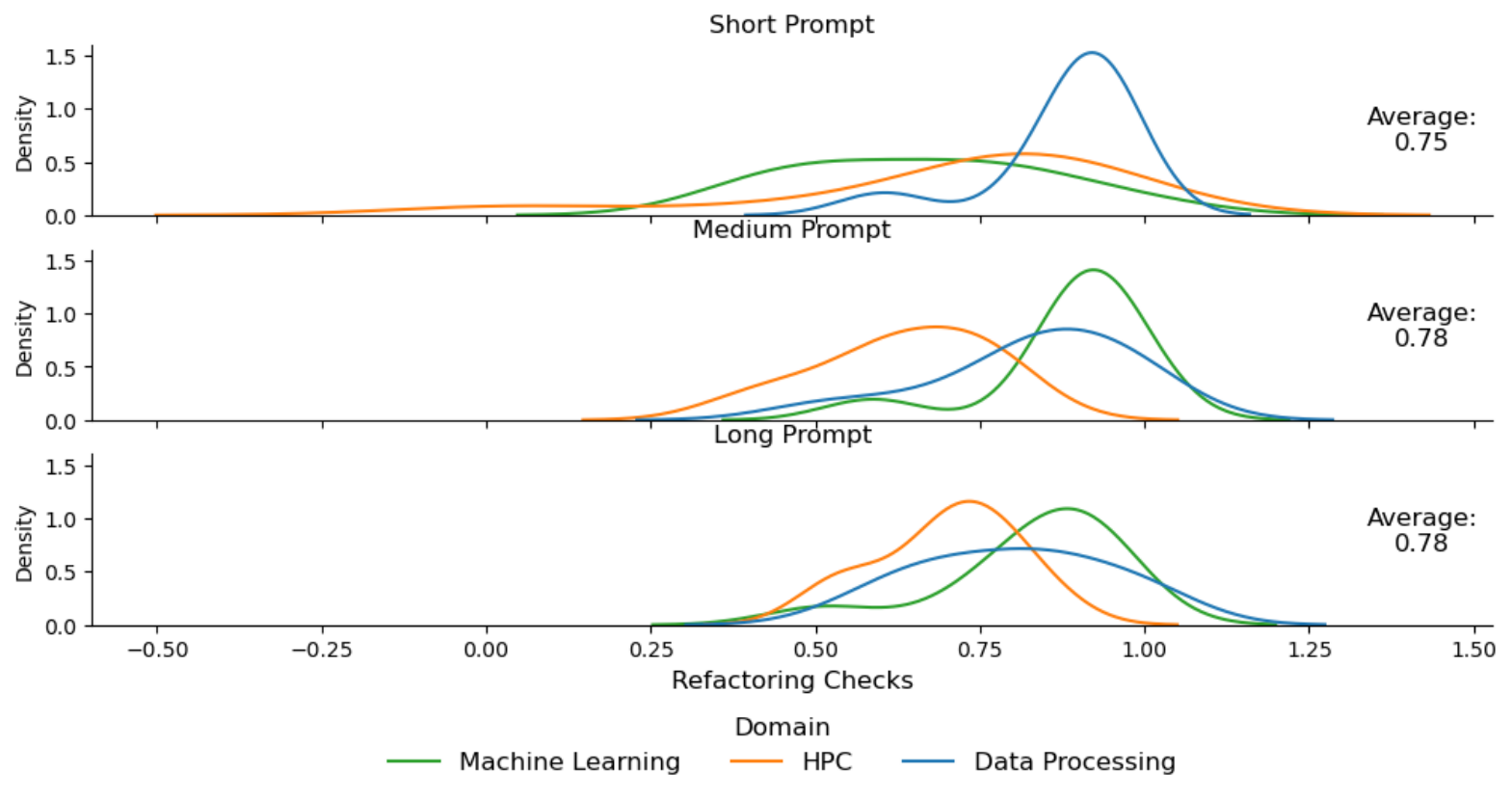}
    \caption{Refactoring checks across three use case}
    \label{fig:domain_refactor}
\end{figure}
Figure~\ref{fig:code_style_violations} shows the scores for the style convention violations of the LLMs. Overall, the LLMs performed significantly poorer in terms of code convention for the short-prompted solution, with a generally higher dispersion of scores and an overall lower average. This dispersion was dramatically reduced in the other prompts, where the distribution was observed to be more compact. This suggests that the specifications for following good Python coding conventions, which are included in these prompts, may have achieved the desired effect. Specifically, the HPC use case is slightly behind in adhering to short and long prompts, which will be addressed later in this section.

Figure~\ref{fig:domain_refactor} shows that the LLMs generally perform better (for each prompt length) in refactoring checks. For the ML domain, the performance was particularly strong for medium and long prompts; however, the LLMs appeared to struggle more with shorter prompts. In contrast, LLMs excel with shorter prompts in the data processing domain. However, the performance became more variable and inconsistent as the length of the prompt increased, which indicates a general lack of stability for solutions with different prompt lengths.

\begin{figure}[!hbt]
\centering
\includegraphics[width=\columnwidth]{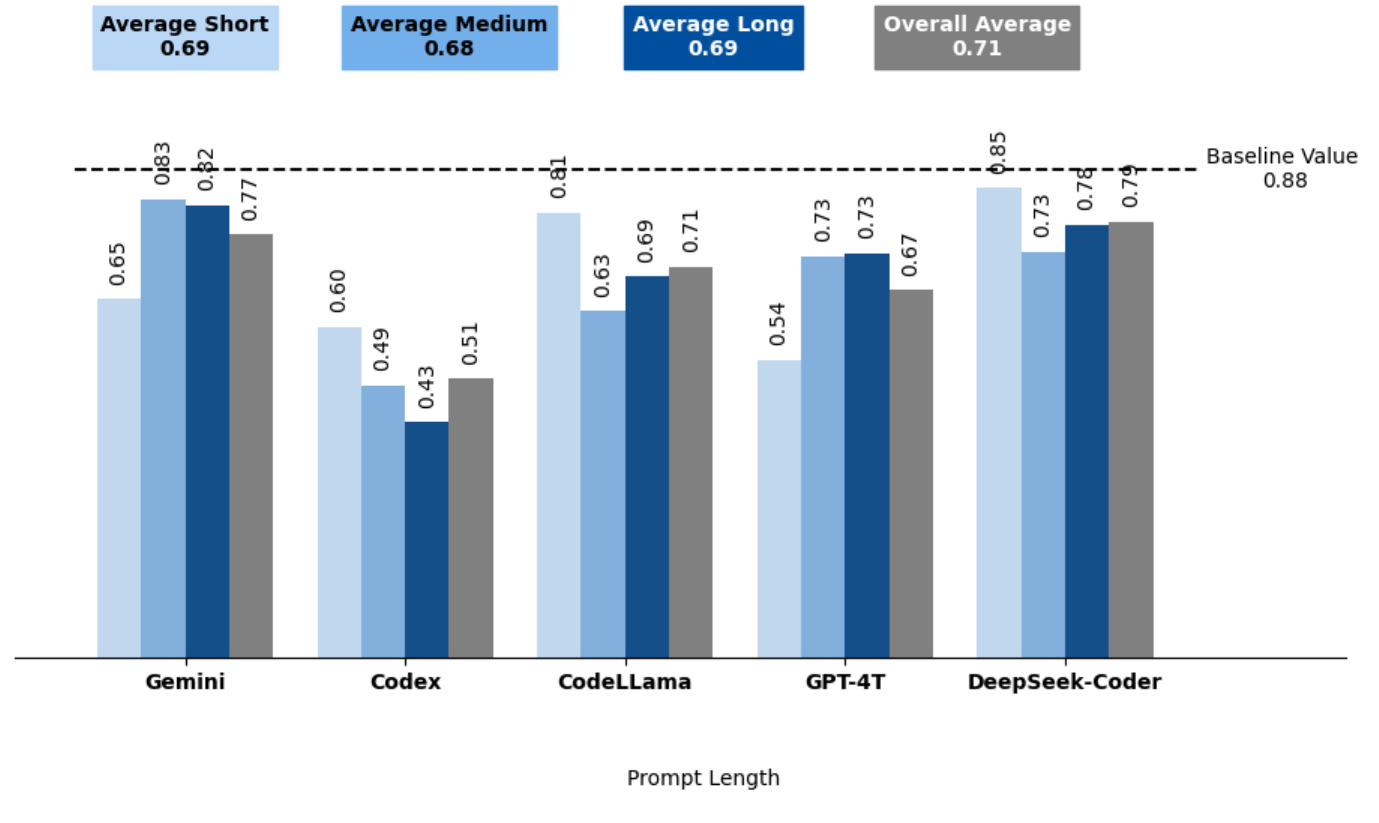}
\caption{Python refactor violations across use cases}
\label{fig:convention_refactoring}
\end{figure}

In Figure~\ref{fig:convention_refactoring}, \texttt{Gemini} compensates for its lower maintainability index performance by achieving the best results in the convention metric, scoring $9\%$ above the average, and ranking second in refactoring checks. As highlighted earlier, the quality of short prompt solutions was generally low regarding convention adherence, with an average score being $23\%$ below the baseline. The lowest performing LLM, \texttt{DeepSeek-Coder}, scores $85\%$ below the baseline for short prompts in convention metrics; however, it excels in refactoring checks, obtaining the best score for short prompts and the highest overall.
\begin{figure}[!hbt]
\centering
\includegraphics[width=\columnwidth]{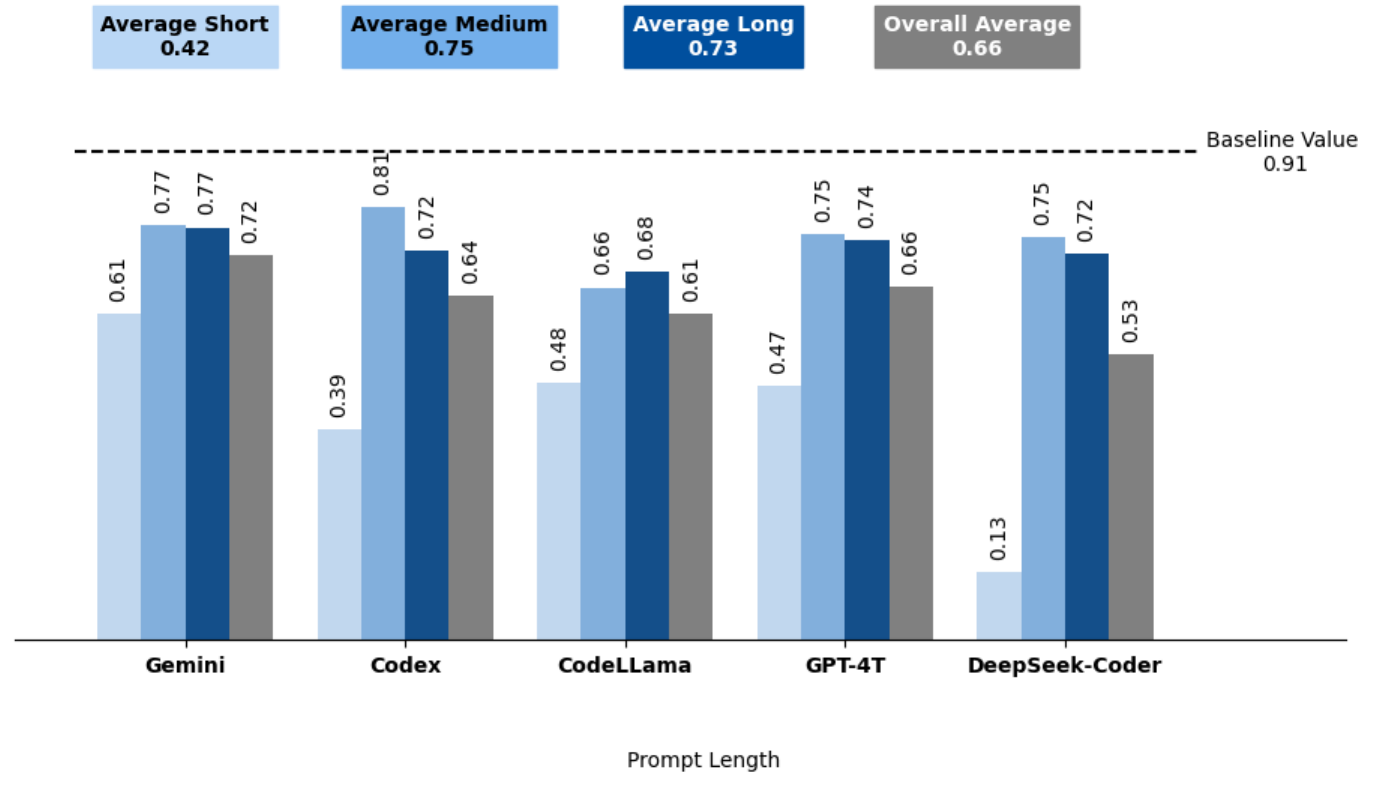}%
\caption{Python style violations across use cases}
\label{fig:convention_domains}
\end{figure}
For refactoring, \texttt{Codex} is the largest underperformer, with an average score of $0.51$, i.e., $28\%$ below other LLMs and $45\%$ below the baseline. No LLM surpasses the baseline for either metric. The convention and refactoring findings suggest that although LLMs can handle the theoretical aspects of balancing complexity, code density, and comment ratios, they struggle to apply these principles within specific programing languages due to language-dependent nuances. The consistently higher performance in refactoring checks may also indicate a lower emphasis on style conventions, which often have less severe implications than other coding errors. This practice could be common and thus more prevalent among the open-source data that many LLMs are trained on. Given that these LLMs \textit{do not adequately manage style conventions or reduce the need for refactoring}, it implies that, depending on the programing language, developers' manual oversight may be crucial to ensure adherence to language-specific best practices.

The overall poorer performance and inconsistencies across prompts for style convention scores (refer to Figure~\ref{fig:convention_domains}) can be attributed to the Strassen algorithm. The Strassen algorithm approaches matrix multiplication as a recursive problem, often reformulating the input matrices into block matrices composed of several smaller matrices, as depicted in Equation~\ref{eq:strassen-notation}~\cite{Zadeh_Santucci_2016}.
\begin{figure}[ht]
    \begin{equation}
        A = \begin{pmatrix}
            A_{11} & A_{12} \\
            A_{21} & A_{22}
        \end{pmatrix},
        \quad
        {B} = 
            \begin{pmatrix} 
                B_{11} & B_{12} \\
                B_{21} & B_{22}
            \end{pmatrix}    
        \label{eq:strassen-notation}
        \end{equation}
    \centering{Equation: Common Strassen Notation}
    \label{fig:strassen-notation}
\end{figure}

\begin{figure}[ht]
    \includegraphics[width=\columnwidth]{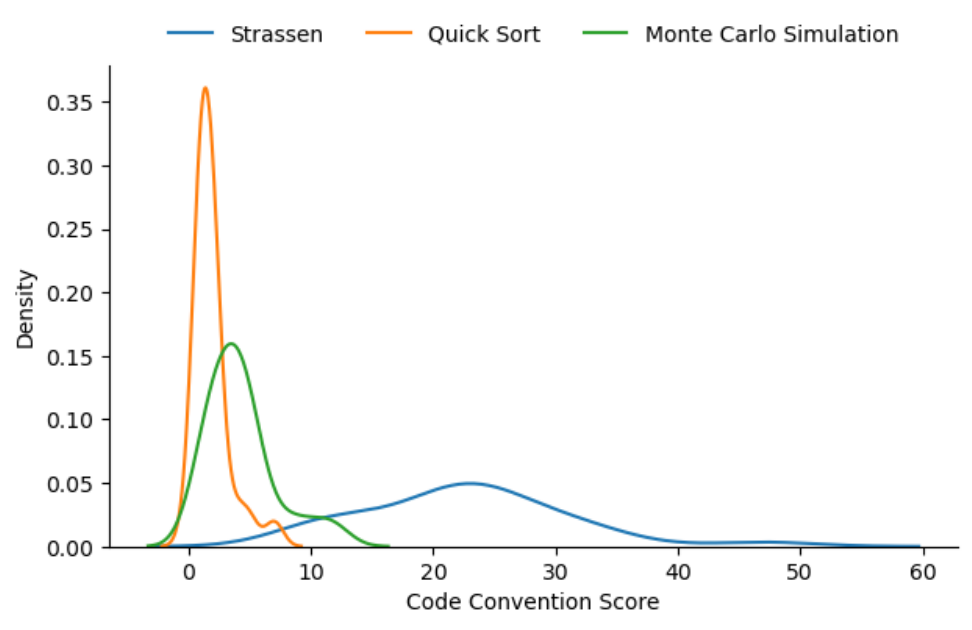}
    \centering
    \caption{Code style convention score distribution}
    \label{fig:hpc_code_convention_density}
\end{figure}

Despite its common use, this notation violates the PEP 8 style guide for variable naming\footnote{\url{https://peps.python.org/pep-0008/\#naming-conventions}}, a discrepancy identified by Pylint. Figure~\ref{fig:hpc_code_convention_density} illustrates the raw count distributions of the style convention violations for the different algorithms within the HPC domain, where the solutions for the Strassen algorithm are significantly more dispersed, with a significantly higher mean, exhibiting $2x$ to $4x$ more code style violations compared to the other algorithms. Although there could be arguments for disabling such warnings in this specific context, LLMs should ideally avoid using improper variable naming and employ more descriptive and informative variable naming practices that comply with Python's PEP 8 style guide.
%
%%%%%%%%%%%%%%%%%%%%%%%%%%%%%%%
\subsubsection{Source Code Statistics}
\label{sec:source_code_stats}
\begin{figure}[!hbt]
    \centering
    \includegraphics[width=\columnwidth]{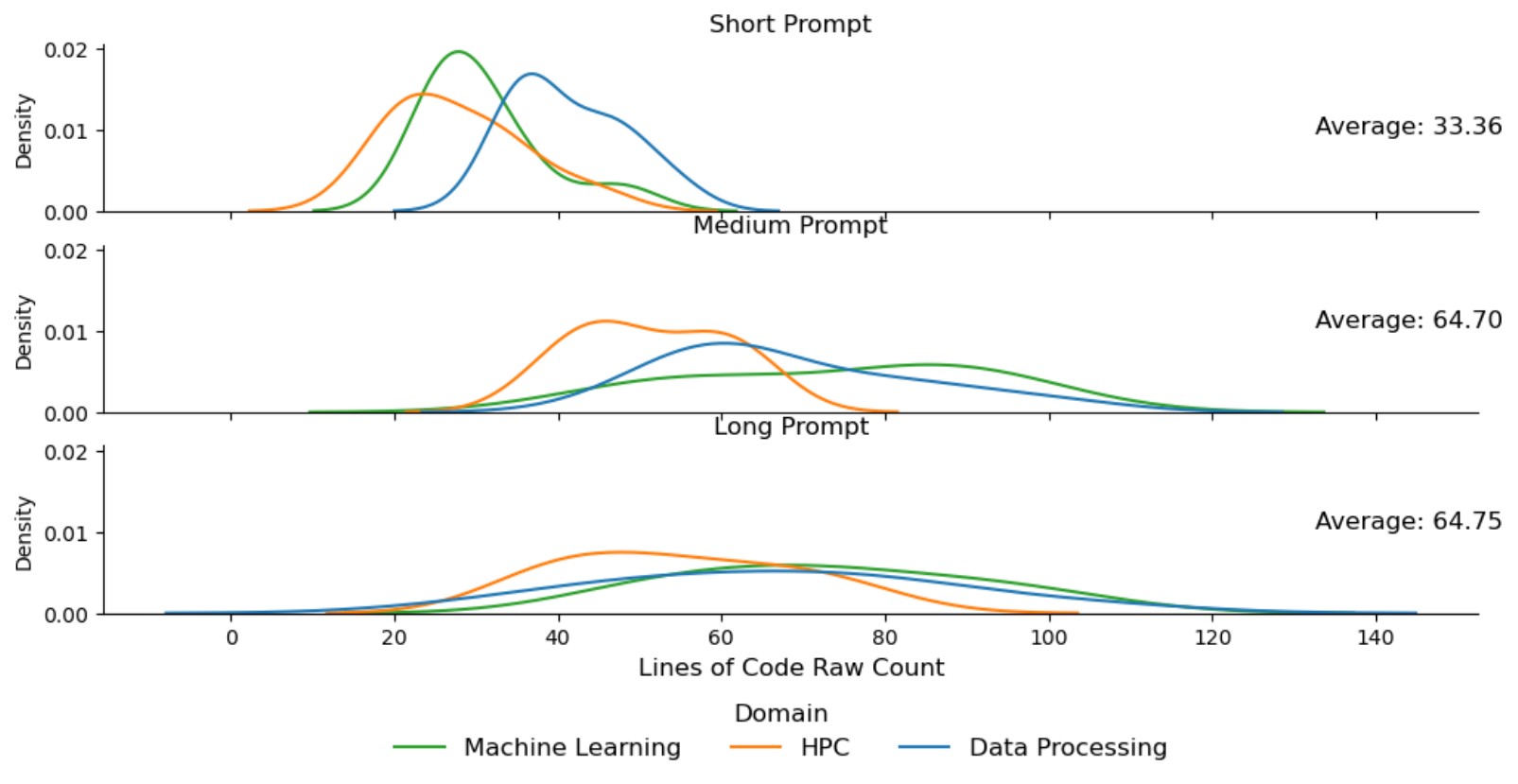}
    \caption{LOC prompt and application domain distributions}
    \label{fig:loc}
\end{figure}

%%%%%%%%%%%%%%%%%%%%%%%%%%%%%%%
Moving on to source code statistics, a clear correlation between increased prompt detail and the length of code produced by the LLMs can be seen (see Figure~\ref{fig:loc}). Medium and long prompt solutions are nearly twice as long as short prompts, indicating that LLMs tend to generate more comprehensive solutions based on detailed requirements specified in the prompts. Additionally, as the prompt length increases, so does the dispersion in the solutions, with shorter prompts typically yielding denser codes and longer ones exhibiting greater variability. This trend implies that as prompts become more detailed, variability in potential solutions also increases, reflecting LLMs’ adaptive responses to expanding requirements.
\begin{figure}[!hbt]
    \centering
    \includegraphics[width=\columnwidth]{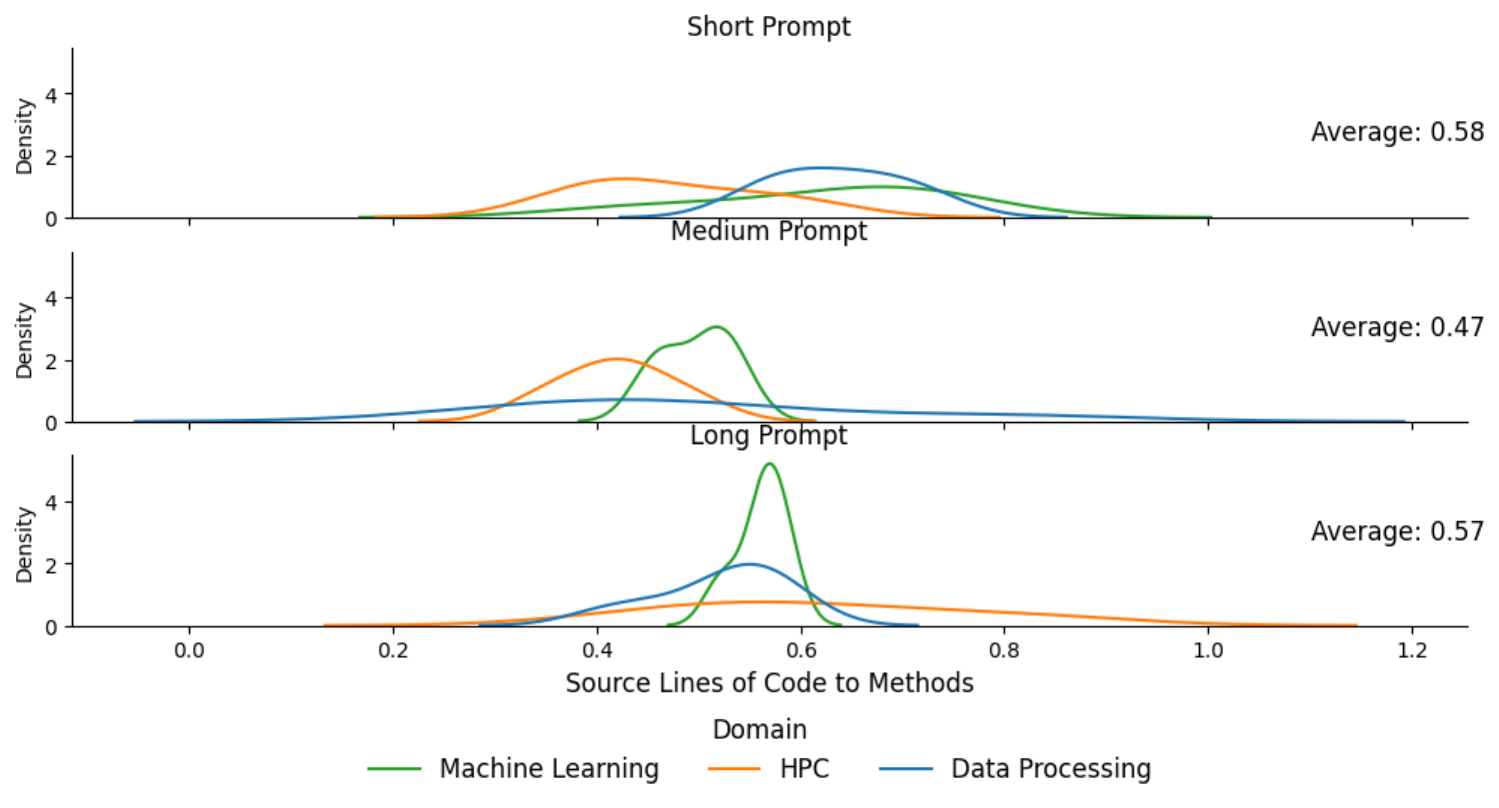}
    \caption{SLOC-to-Methods prompt and application domain
distributions}
    \label{fig:sloc_to_methods_domain}
\end{figure}
The ability of LLMs to disperse logic across multiple methods shows significant variability, as indicated by the wide dispersion in solutions depicted in Figure~\ref{fig:sloc_to_methods_domain}. Notable variations are observed for the data processing domain with medium prompts, HPC for long prompts, and ML for short prompts, highlighting domain-specific differences. Despite similar details in the prompts, the distributions varied considerably, with medium prompts performing $18\%$ lower on average than long prompts. Each domain exhibited a more consistent dispersion around the mean for certain prompt lengths, suggesting some uniformity within this variability.

The significant dispersion and minimal correlation across domains underscore the domain-specific differences previously noted in Python code-style violations (section~\ref{sec:Python Coding Violations}). Additionally, the variability within domains suggests the challenge of developing a universal prompting methodology that effectively spans different domains, reflecting each domain's complexity and unique requirements in terms of how LLMs generate and structure code.

\begin{figure}[ht]
    \centering
    \includegraphics[width=1\columnwidth]{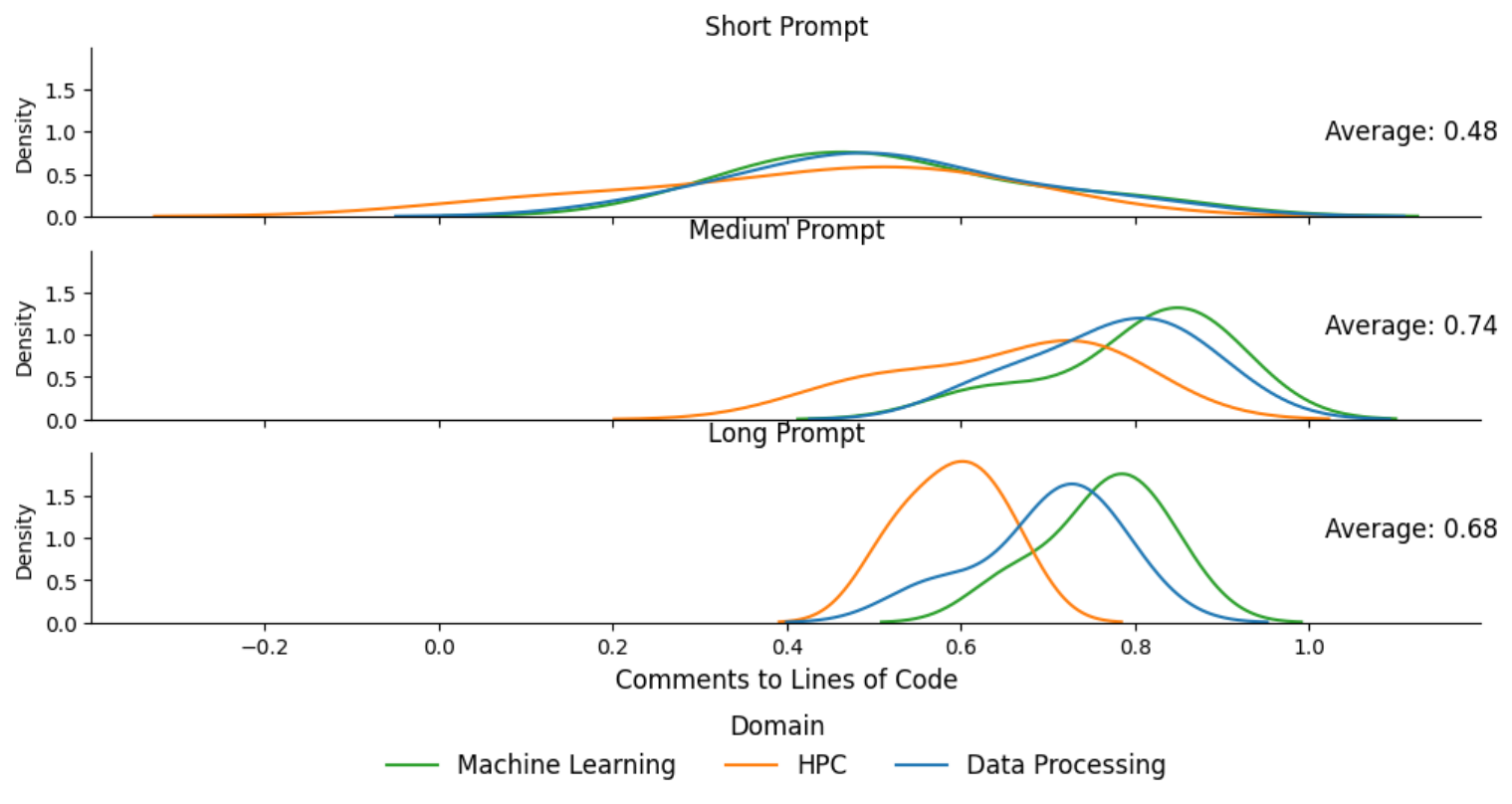}
    \caption{Comments-to-code ratio with varying prompts and application domain
distributions}
    \label{fig:c_to_loc}
\end{figure}

\begin{figure}[ht]
    \includegraphics[width=\columnwidth]{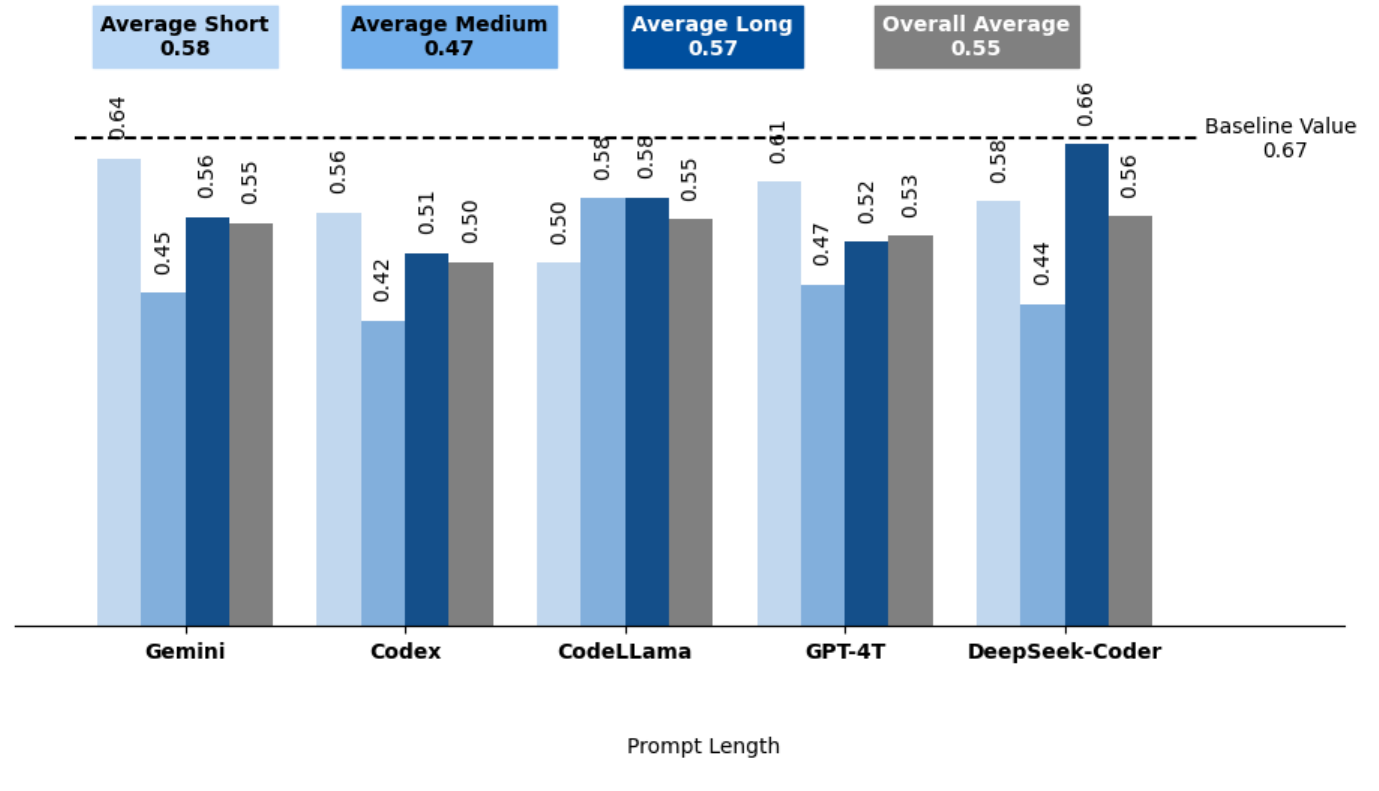}
     \caption{SLOC-to-methods by LLMs}
\label{fig:code_statistics_maintainability}     
\end{figure}

Despite the overall variations, the comment-to-code ratio, which includes docstrings, shows more similarity across medium and long prompts than short ones (refer to Figure~\ref{fig:c_to_loc}). The longer prompts, which emphasize proper commenting, have higher averages than the short prompt solutions, which was intentionally requested within these prompts, indicating better LLM adherence when requested. In particular, medium prompts unexpectedly obtained slightly higher averages than long prompts despite having similar solution lengths (see Figure~\ref{fig:c_to_loc}). This suggests that LLMs effectively integrate comprehensive documentation per request of the prompt, efficiently producing thorough solutions even without requiring extra detail from long prompts. Conversely, short prompts tend to yield shorter, less documented solutions.
Generally, the source lines of code (SLOC)-to-methods ratio and the comment-to-code ratio exhibit wider dispersion (refer Figure~\ref{fig:code_statistics_maintainability} and Figure~\ref{fig:c_to_loc}), indicating limited similarity in these metrics across different domains and prompt lengths. However, the deliberate inclusion of requests for commenting and docstrings in the medium and long prompts appears to have had the intended effect, successfully nudging LLMs toward improved maintainability.

\begin{figure}[ht]
    \centering
    \includegraphics[width=\columnwidth]{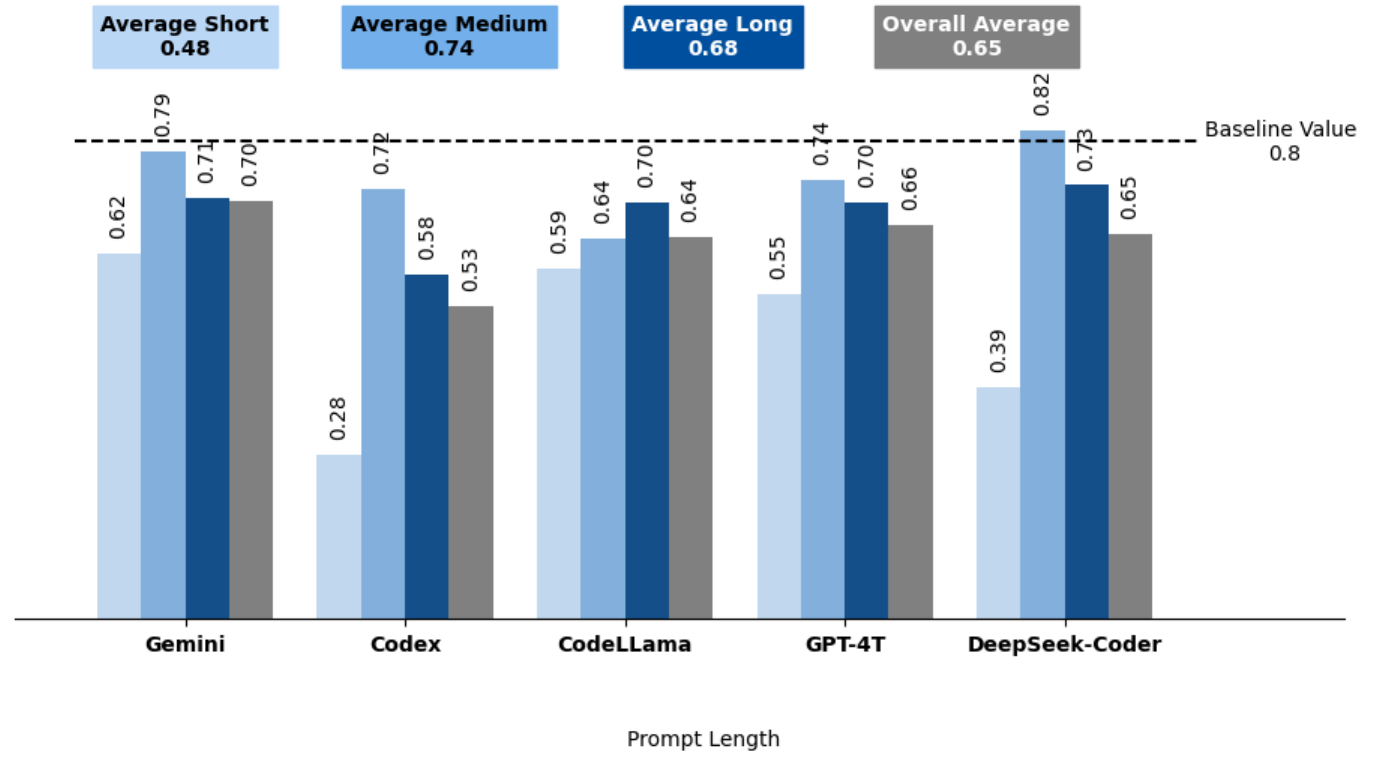}%
  \caption{Comments-to-code ratio by LLMs}
\label{fig:comment-to-code}
\end{figure}

On average, no LLM significantly outperforms others for SLOC per method (shown in Figure~\ref{fig:code_statistics_maintainability}). Different LLMs excel with specific prompt lengths, with \texttt{Gemini} leading for short prompts, \texttt{CodeLLama} for medium, and both \texttt{DeepSeek-Coder} for long prompts. However, their higher performance with certain prompts is balanced by less notable results within others. Notably, no LLM is capable of matching the high score achieved by the baseline.
For the comment-to-code ratios (refer to Figure~\ref{fig:comment-to-code})), \texttt{Gemini} on average, produced the best solutions, followed by \texttt{GPT-4T}, \texttt{DeepSeek-Coder} and \texttt{CodeLLmama}. However, no LLM outperformed the baseline, with the average of \texttt{Gemini} being $13\%$ below the baseline. Similar to the SLOC method, \texttt{DeepSeek-Coder} produces a solution set for a single prompt with higher code documentation.
In particular, for neither of the metrics, no LLM is capable, on average, of producing solutions that outperform the established baseline, with only two sets of solutions overall achieving higher scores. This may reflect the natural differences in coding styles compared to human-made baseline solutions, where common and widely used best-practice coding standards may be embedded more into the developer style. Compared to LLMs, they may be more influenced by the nature of the open-source code frequently used to train LLMs, which may exhibit less maintainable characteristics in the source code, resulting in slightly lower scores. A similar argument was made earlier while discussing the differences between convention violation and refactoring checks.

%%%%%%%%%%%%%%%%%%%%%%%%%%%%%%%%%%%%%%%%%%%%%%%%%%%%%%%%%%%%%%%%%%%%
\subsection{Performance}
%%%%%%%%%%%%%%%%%%%%%%%%%%%%%%%%%%%%%%%%%%%%%%%%%%%%%%%%%%%%%%%%%%%%
Preprocessing the CPU metric only by standardizing it to the range of zero to one by dividing by 100. The memory usage was contextually normalized, which improved the reliability of the comparison.

%%%%%%%%%%%%%%%%%%%%%%%%%%%%%%%%%%%%
\subsubsection{CPU Usage}
\label{sec:cpu_usage}
\begin{figure}[ht]
    \centering
    \includegraphics[width=\columnwidth]{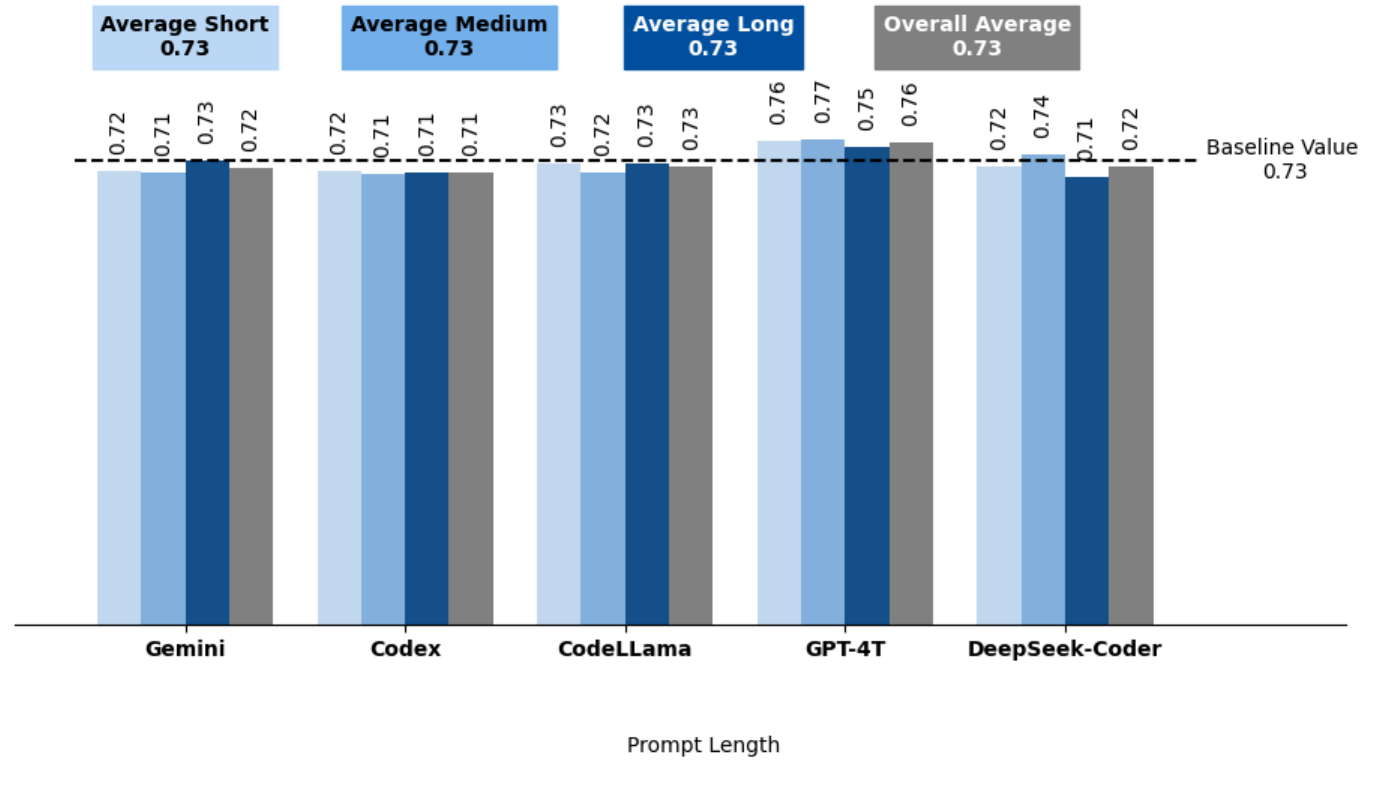}
    \caption{CPU Usage by all five LLMs}
    \label{fig:cpu_hpc}
\end{figure}

%%%%%%%%%%%%%%%%%%%%%%%%%%%%%%%%%%%%
Figure~\ref{fig:cpu_hpc} shows that all LLMs use comparable CPU resources. Overall, \texttt{GPT-4T} achieve the highest average score and the most efficient solutions for all prompts. However, these differences represent only an average increase of $4\%$ over the baseline.
%

%%%%%%%%%%%%%%%%%%%%%%%%%%%%%%%%%%%%
\subsubsection{Memory Usage}
\label{sec:mem_usage}
\begin{figure}[!hbt]
    \centering
    \includegraphics[width=\columnwidth]{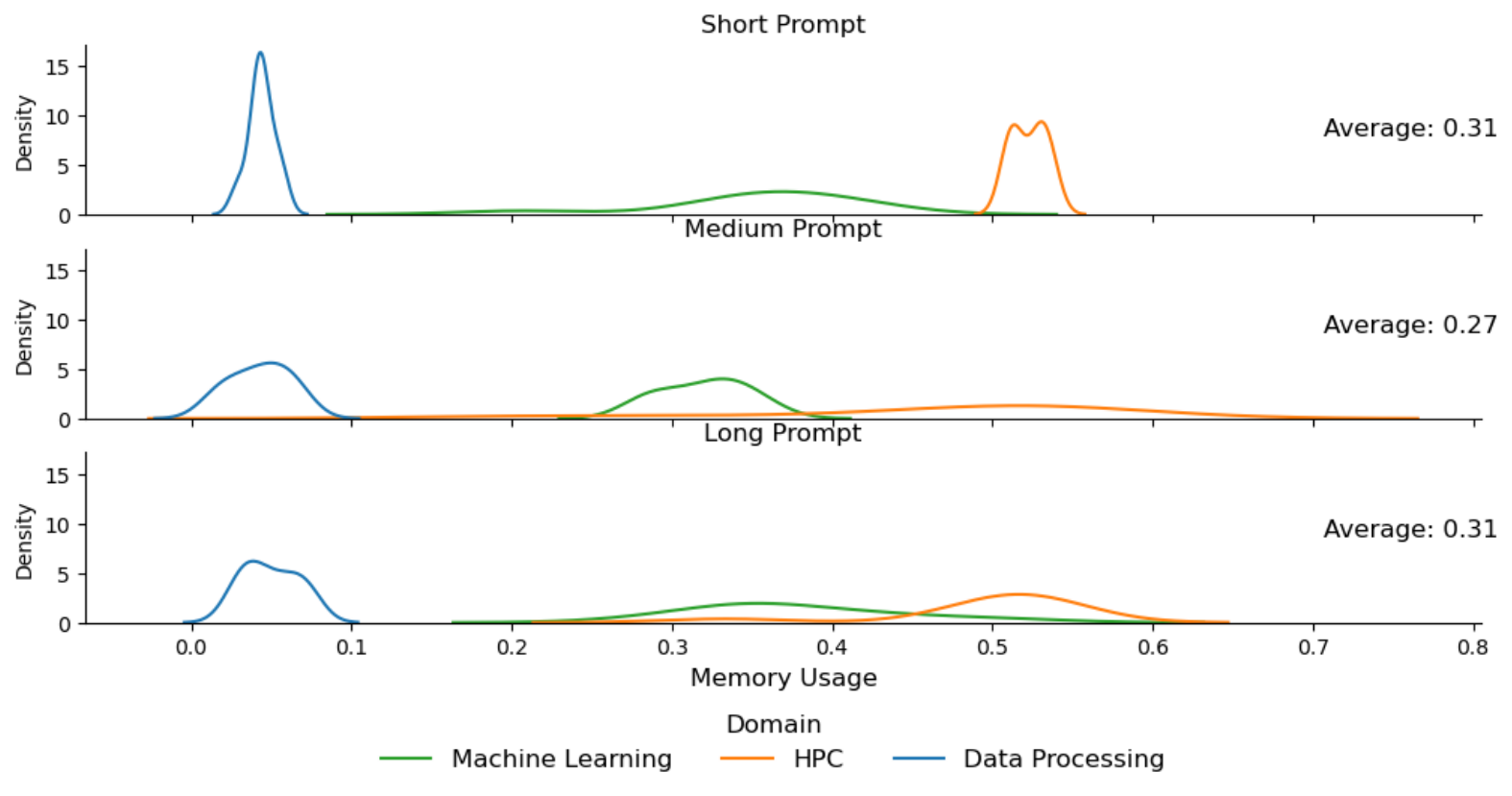}
    \caption{Memory usage across prompts and application
domain distributions}
    \label{fig:mem_usage_dist}
\end{figure}

%%%%%%%%%%%%%%%%%%%%%%%%%%%%%%%%%%%%
Exploring domain memory usage, as illustrated in Figure~\ref{fig:mem_usage_dist}, there is a significant variation between distributions. The ML and HPC domains exhibit higher variability in solutions for memory usage, whereas the data processing domain has a very high concentration of similar results. The maximum observed value for each algorithm was used to contextually normalize memory usage, and the metric was inverted to assign higher scores to lower values. Therefore, the low values in the data processing domain indicate that most values are close to the maximum, suggesting consistent memory usage across different prompts and LLMs. Conversely, other domains, especially outside of short prompt solutions in HPC, show much lower density, indicating greater variability between high and low memory usage in the provided solutions.
\begin{figure}[!hbt]
    \centering
    \includegraphics[width=\columnwidth]{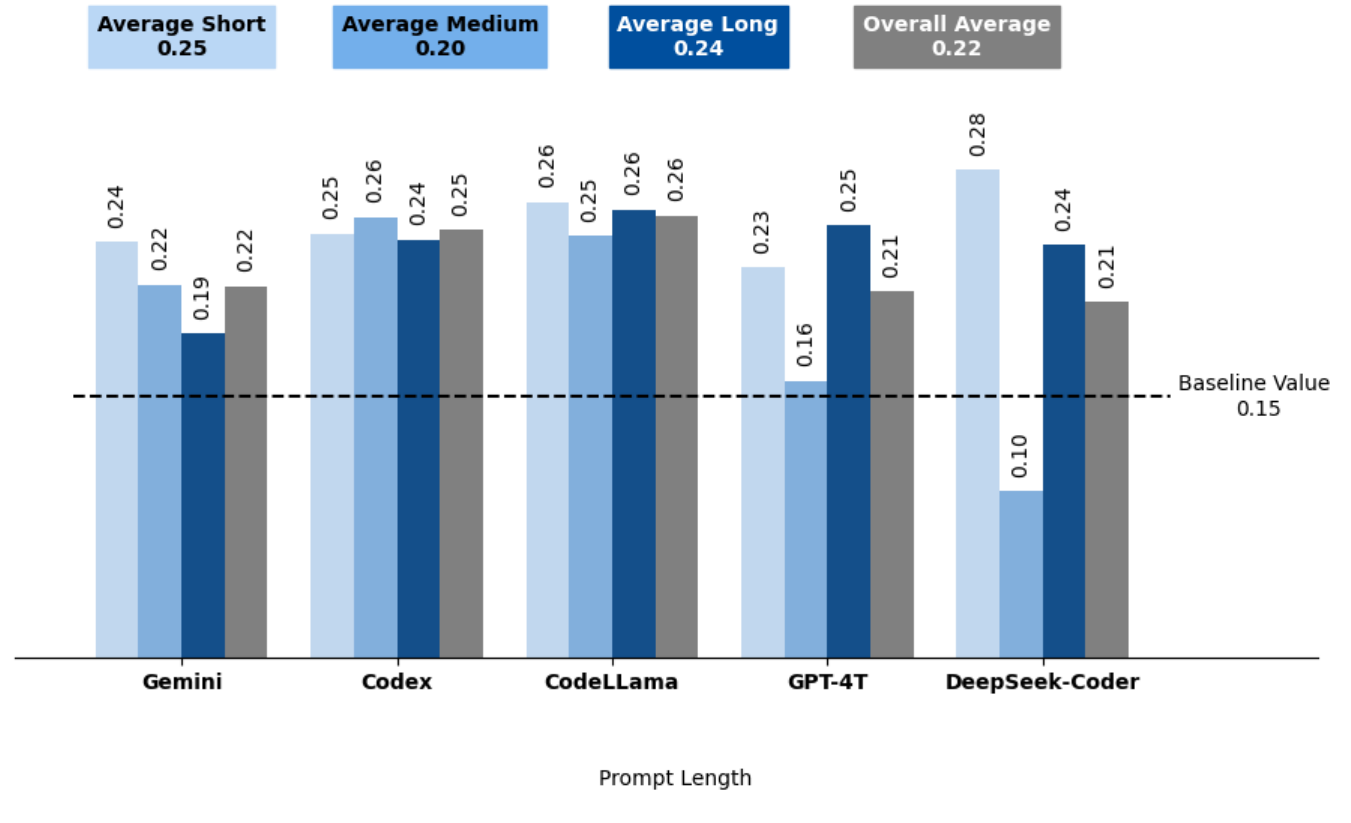}
    \caption{Memory usage by all five LLMs}
    \label{fig:mem_usage_model}
\end{figure}
We can see that (refer Figure~\ref{fig:mem_usage_model}), \texttt{DeepSeek-Coder} shows almost a $4x$ difference in memory usage between its short and medium prompt solutions. Similarly, \texttt{GPT-4T} and \texttt{Gemini} also exhibit big variations between their highest and lowest performance. In contrast, LLMs like \texttt{Codex} and \texttt{CodeLLama} demonstrate exceptional consistency across prompts, ranking among the best overall. \texttt{DeepSeek-Coder}'s medium prompt solution failed to outperform the baseline, and all other solutions surpassing it.

We found that LLMs with the worst memory usage matched LLMs with the best maintainability. In contrast, LLMs that demonstrate high variability in memory usage, such as \texttt{GPT-4T} and \texttt{Gemini} rank among the highest in terms of maintainability. This inverse relationship highlights differing strengths and trade-offs in performance and highlights the importance of considering their inherent strengths and weaknesses before employing them in practice.

%%%%%%%%%%%%%%%%%%%%%%%%%%%%%%%%%%%
\subsection{Reliability}
%%%%%%%%%%%%%%%%%%%%%%%%%%%%%%%%%%%%
Metrics, such as cyclomatic complexity and Halstead's number of delivered bugs, provide insights into the general reliability of the source code. Here, Python-specific metrics are used (using Pylint), which provides warnings and error messages. If these are left unattended, the resulting code can lead to unexpected code behavior, thereby impacting operational reliability.

%%%%%%%%%%%%%%%%%%%%%%%%%%%%%%%%%%%%
\subsubsection{Complexity and Bugs}
\label{sec:bugs_and_complexity}
\begin{figure}[!hbt]
    \centering
    \includegraphics[width=\columnwidth]{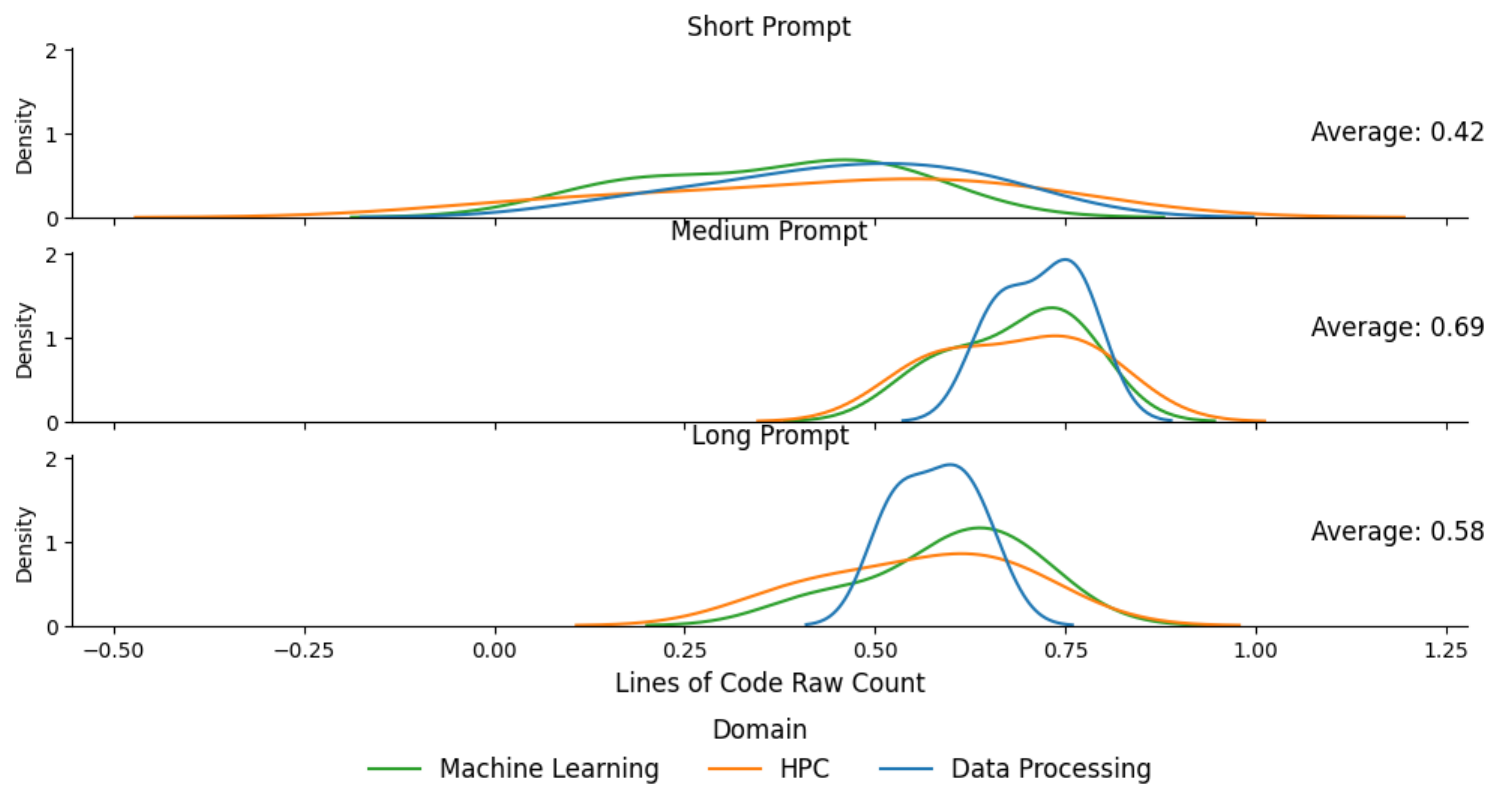}
    \caption{Complexity across prompt and application domain}
    \label{fig:complexity_distribution}
\end{figure}

%%%%%%%%%%%%%%%%%%%%%%%%%%%%%%%%%%%%
The LLMs exhibit less consistency across prompts when evaluating the complexity of the methods shown in Figure~\ref{fig:complexity_distribution}. Here, solutions tend to share more similarities across domains within the same prompt than across prompts in the same domain. This may suggest that LLMs opt for similar complexity densities within prompts, for example, aggregating all logic and decisions within one method in a small prompt but opting for additional methods in longer prompts. Given the predefined complexity of the algorithms, this naturally reduces the complexity of longer prompts. An interesting observation is that despite similar code length (as seen in Section~\ref{sec:source_code_stats}), the longer prompt solutions performed slightly better than medium ones, which suggests more dense complexity and less dispersion. One reason for this could be longer prompts introducing over-specifications, which would force LLMs to generate a more elaborate code than necessary, potentially adding unnecessarily increasing complexity.
\begin{figure}[!hbt]
    \centering
    \includegraphics[width=\columnwidth]{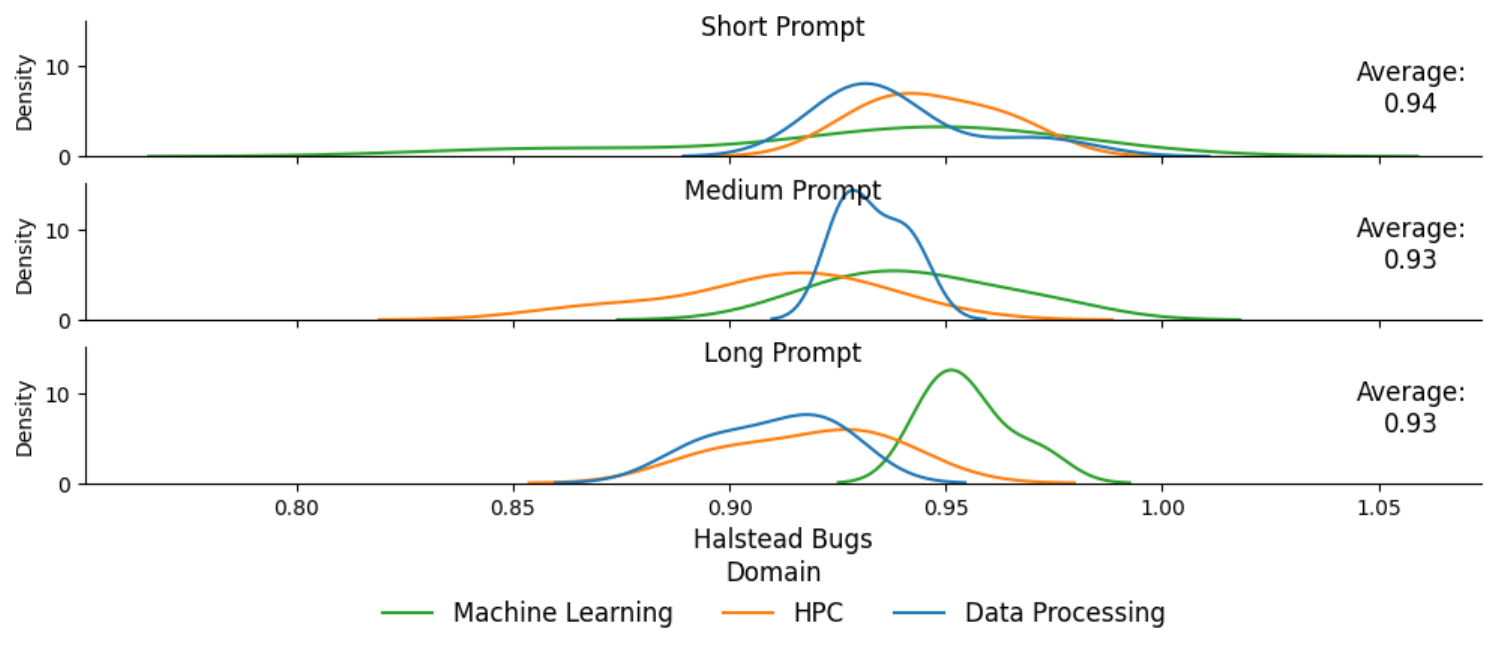}
    \caption{Halstead bugs across prompt and application
domain}
    \label{fig:halstead_bugs}
\end{figure}
LLM generally provides more consistent results relative to the number of delivered bugs across domains within the same prompt (see Figure~\ref{fig:halstead_bugs}) than across prompts. In addition, short-prompt solutions tend to deliver the highest number of bugs. Considering that the delivered bugs are calculated using operands and operators combined with the precise specifications required by the algorithms, we expect a similar score across the prompts.
\begin{figure}[!hbt]
   \centering
    \includegraphics[width=\columnwidth]{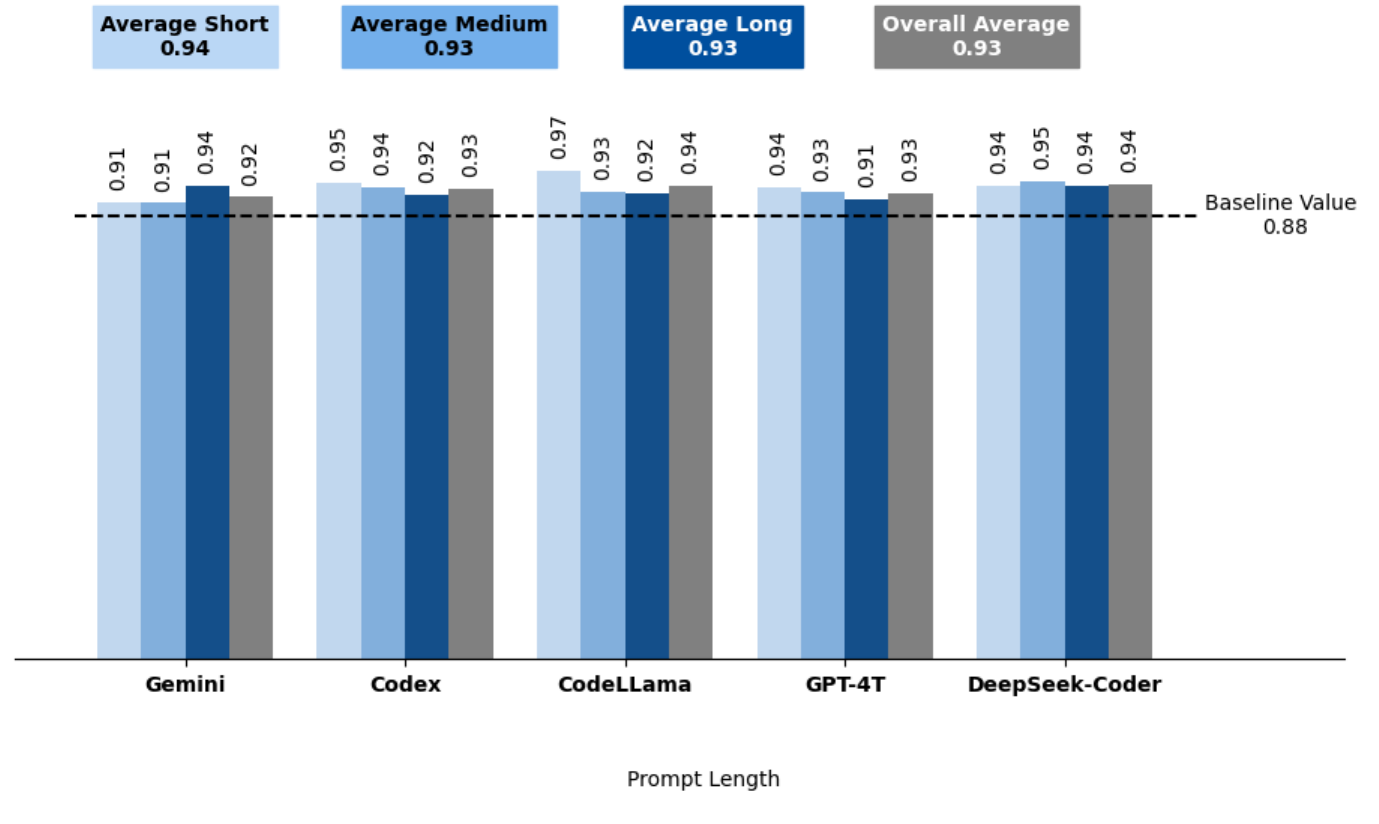}%
    \caption{Number of delivered bugs by each LLMs}
    \label{fig:bugs_}
\end{figure}
\begin{figure}[!hbt]
   \centering
   \includegraphics[width=\columnwidth]{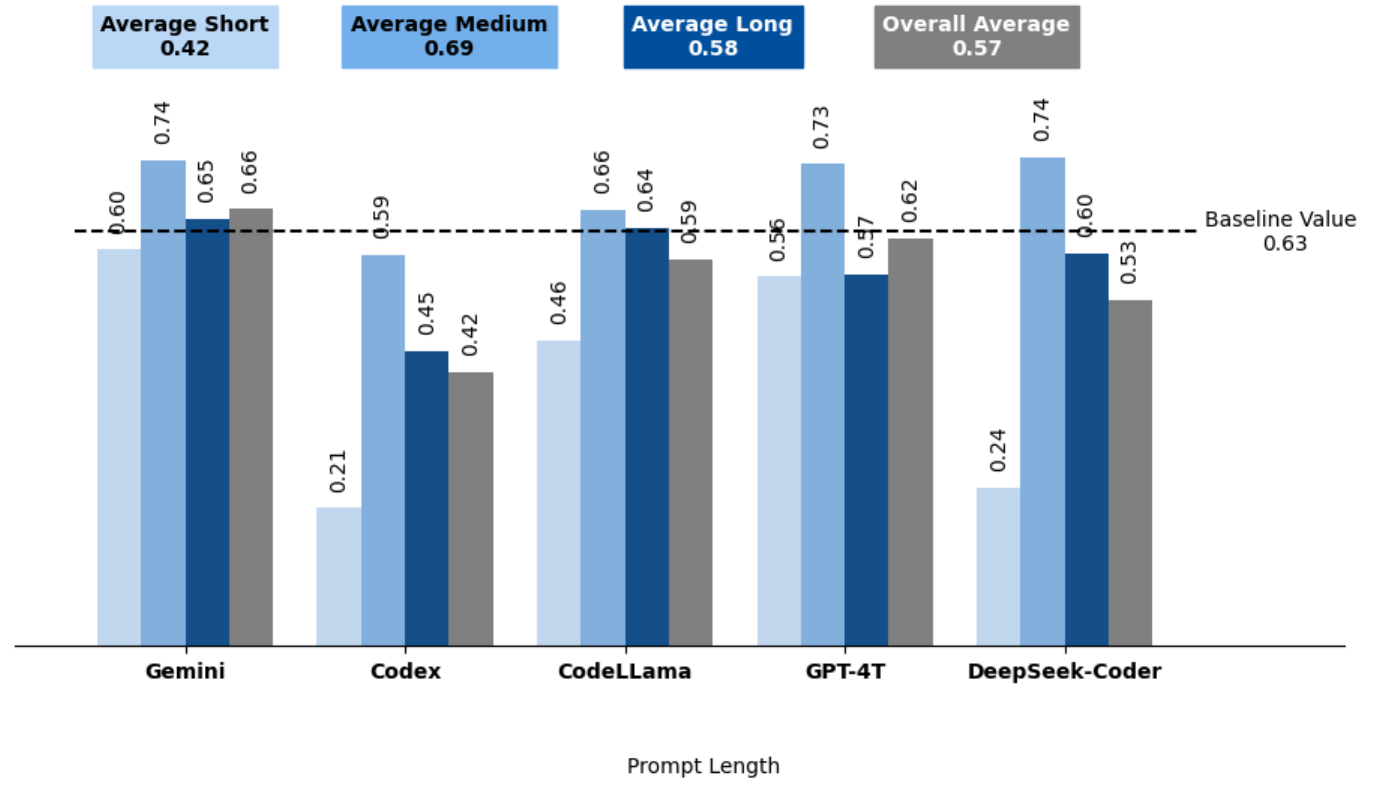}%
  \caption{Cyclomatic complexity by each LLMs}
\label{fig:_complexity}
\end{figure}

In analyzing the two metrics across each LLM shown in Figure~\ref{fig:bugs_}~and in Figure~\ref{fig:_complexity}, it is clear that the consistency is general for all LLMs given the similarity in score with no significant deviations observed. Notably, \texttt{CodeLLama} records the fewest delivered bugs in its short prompt solutions (refer to Figure~\ref{fig:bugs_}), though this advantage diminishes when averaged across all prompt lengths. A key observation is the underperformance of the baseline model, which scored lower than all models. Given that bugs are computed from operands and operators, this suggests that baseline solutions include an unnecessary amount. In contrast, LLMs adhere more closely to the specified requirements in the prompts and reduce the number of performed operations to what is strictly necessary.

This consistency is far less so for complexity, with more significant changes and variations appearing, suggesting different approaches to dispersing complexity in the code, as also seen for domain distributions. In particular, for short prompts, there was a significantly lower complexity (refer to Figure~\ref{fig:_complexity}) achieved by \texttt{Gemini}, \texttt{GPT-4T} and \texttt{CodeLLama} (a higher score refers to better complexity dispersion). In particular, the two former deviate significantly, producing solutions that achieve higher scores than the prompt average. These LLMs produce similarly low-complexity solutions for medium prompts, and their overall average outperforms other prompts. Considering this considering maintainability, this suggests that these LLMs can produce code that upholds best practices and ensures that the complexity is well disseminated across methods. Similar to the findings from the domain distributions, medium-prompt solutions, on average, are the best, exceeding the established baseline, with many LLMs also managing to outperform it for medium-prompt solutions.

%%%%%%%%%%%%%%%%%%%%%%%%%%%%%%%%%%%%
\subsubsection{Python Specific Reliability}
\label{sec:python_specific_reliability}
%%%%%%%%%%%%%%%%%%%%%%%%%%%%%%%%%%%%
%
\begin{figure}[!hbt]
\centering
\includegraphics[width=\columnwidth]{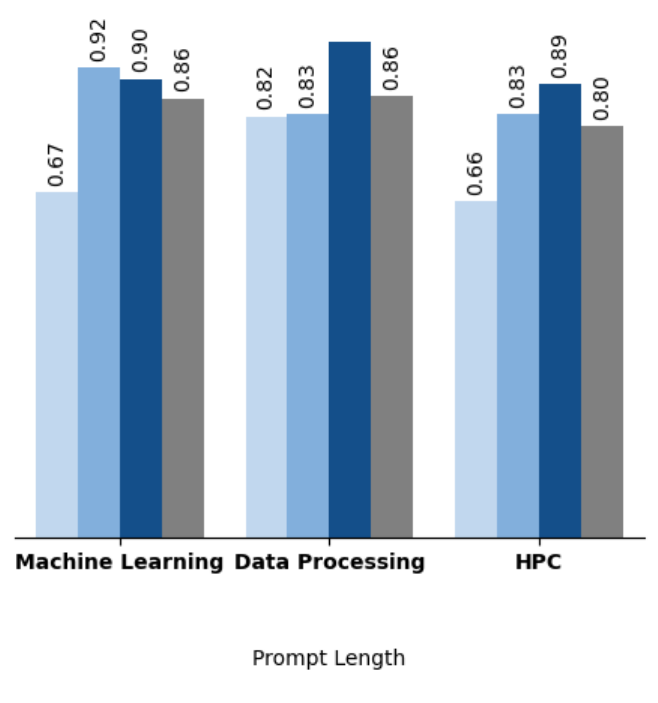}%
\caption{Warning messages across prompts and application
domains}
\label{fig:warnings_and_domain}
\end{figure}
\begin{figure}[!hbt]
\centering
\includegraphics[width=\columnwidth]{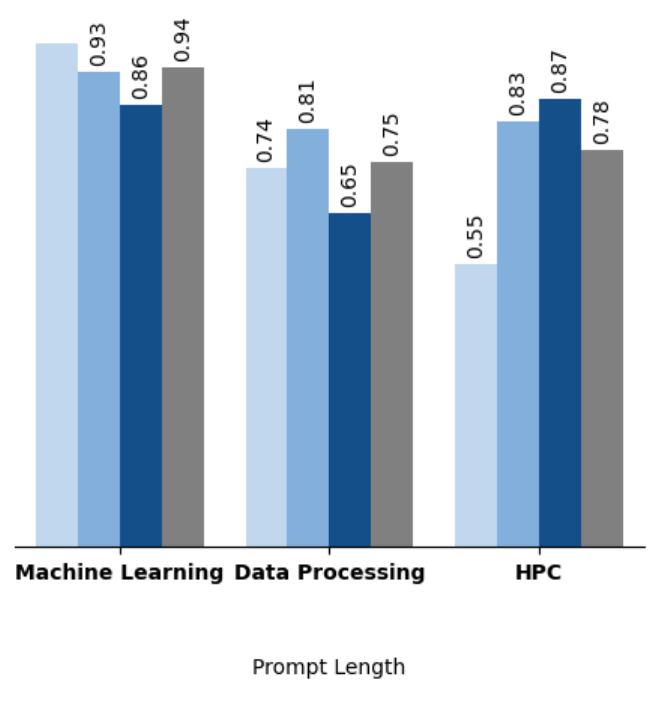}%
\caption{Error messages across prompts and application
domains}
\label{fig:errors_and_domain}
\end{figure}
\begin{figure}[!hbt]
\centering 
\includegraphics[width=\columnwidth]{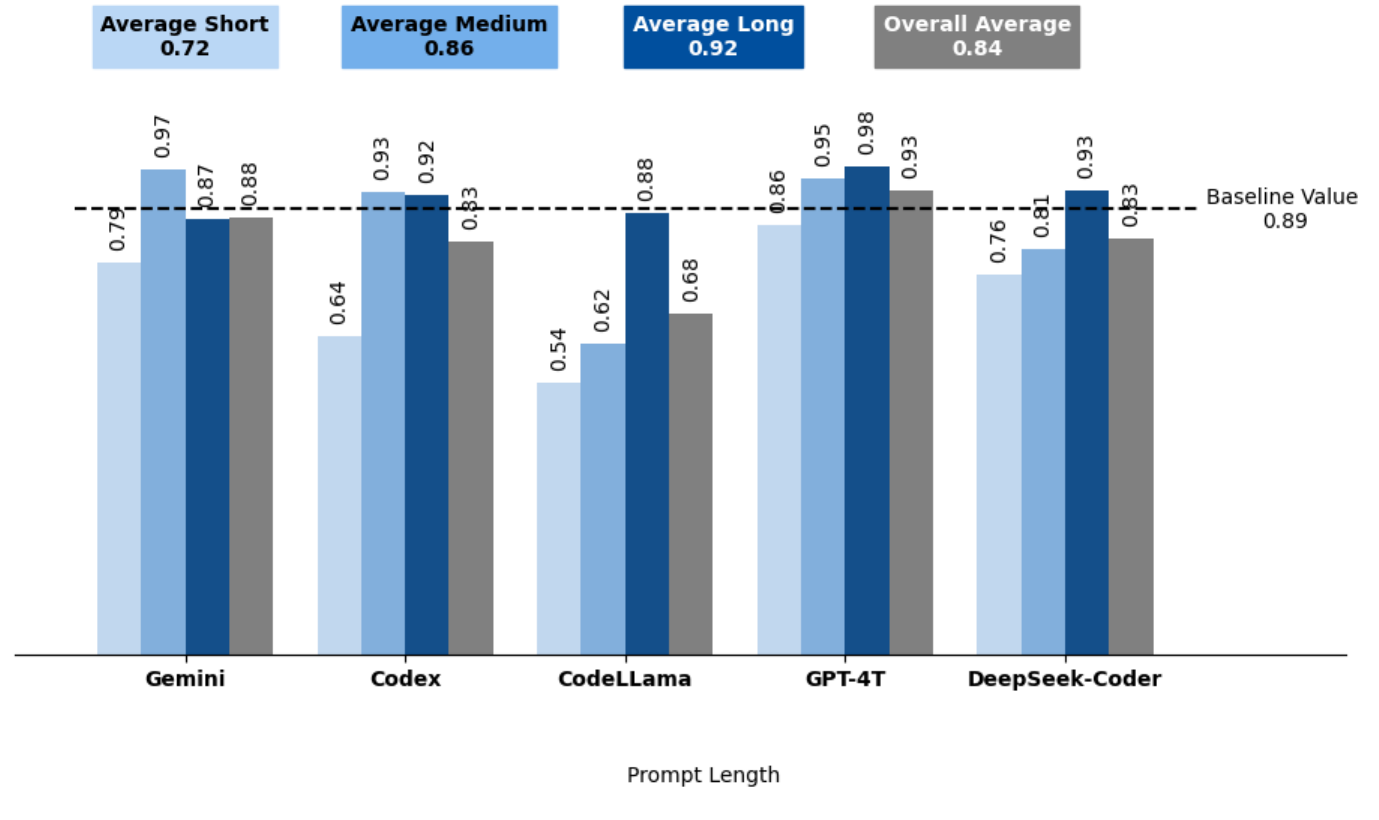}
\caption{Warning messages by each LLMs}
\label{fig:warning_llms}
\end{figure}
\begin{figure}[!hbt]
\centering
\includegraphics[width=\columnwidth]{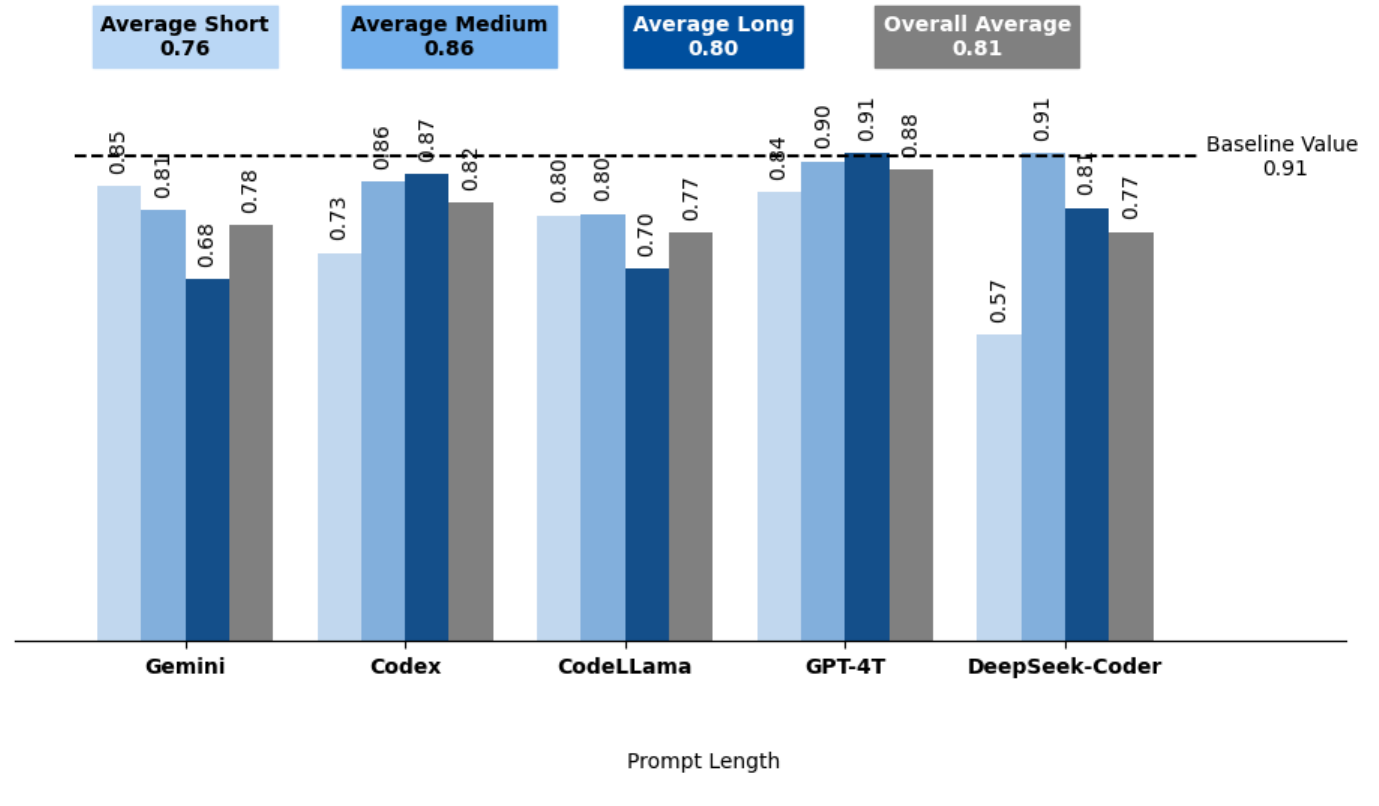}%
\caption{Errors messages by each LLMs}
\label{fig:error_llms}
\end{figure}
When examining the Python-specific reliability metrics across domains in Figure~\ref{fig:warnings_and_domain}, warning messages exhibit little variation, with most domains exhibiting consistent results across prompts and domains. Note that some notable exceptions include the short prompt solution for the ML and HPC domains, where warnings are slightly more frequent.

However, this consistency does not extend to errors that exhibit variability within and between domains (refer to Figure~\ref{fig:errors_and_domain}. In particular, solutions in the HPC and Data Processing domain display poorer quality and more errors than ML. Results also significantly differ across prompts in the same domain. Although these metrics primarily relate to correct Python programing, the unique characteristics of each domain and possibly a lack of suitable training data examples may explain the observed inconsistencies, especially the lower performance observed in HPC.

The LLM performance for Python-specific reliability metrics (refer to Figure~\ref{fig:warning_llms} and Figure~\ref{fig:error_llms}) shows some notable variability. \texttt{CodeLLama} tend to generate more warnings, which deviated significantly from the average values. In contrast, \texttt{GPT-4T} consistently produces the fewest warnings and errors (see Figure~\ref{fig:error_llms}), performing well above the overall average.

Interestingly, although the short prompt solutions obtained the lowest scores, they were not far off the performance achieved for the long prompts. This may indicate that LLMs can generally avoid more significant code violations, with the deviations likely being influenced by LLM-specific characteristics rather than prompt length alone. Given the higher performance exhibited by \texttt{GPT-4T}, this difference is likely caused by the less refined training datasets.

Although the metrics demonstrate variability with some notable underperformers, most LLMs maintain a generally high level of performance. As discussed previously, this may be due to the higher severity of warnings and errors, which would likely occur less frequently in training data than the style-related convention and refactoring metrics seen in Section~\ref{sec:results_maintainability}.

%%%%%%%%%%%%%%%%%%%%%%%%%%%%%%%%%%%%%%%%%%%%%%%%%%%%%%%%%%%%%%%%%%%%%%%%
\section{Discussions and Limitations}
%%%%%%%%%%%%%%%%%%%%%%%%%%%%%%%%%%%%%%%%%%%%%%%%%%%%%%%%%%%%%%%%%%%%%%%%
In the previous sections, we have demonstrated that three main factors influence the effectiveness of LLMs in code synthesis. The most prominent factor was the \textit{prompt design}, which influenced the overall code quality. This contradicts the otherwise intuitive expectation that more details and longer prompts will improve code quality. Next, \textit{application domain}-related differences appeared similarly, and while these varied in significance, their presence highlights inherent inconsistencies that can influence code quality across domains. Each domain demonstrated high and low densities across various metrics, and even within the same metrics across different prompts. This variability highlights the importance of careful consideration and thorough evaluation of LLMs before deployment in unknown contexts, ensuring that their strengths align with specific domain requirements and challenges. Finally, the discrepancy between the updated \textit{versions of LLMs (or variants of LLMs)} highlights the obvious effect of an LLM's lifecycle on its inherent support, maintenance, and overall inference quality. Given these factors, it is clear that developers must be mindful of various factors and that the selected LLM must also be chosen, emphasizing its strengths and weaknesses in the context under consideration. Preferably, after applying the LLM, a thorough evaluation of its inherent strengths and weaknesses in the domain-specific context is performed. By doing so, discrepancies like those found for Python regarding maintainability and reliability can be identified and addressed.

\subsection{Limitations}
This study, while comprehensive, has certain limitations that could influence the generalizability of the findings. First, the selection of LLMs was limited to five state-of-the-art LLMs which could limit the broader applicability of the results. Second, this study focused on real-world programing challenges to reflect various contexts in which LLMs are likely to be used. However, this approach may not fully capture the complex and unpredictable nature of a problem, which often involves more diverse and complex requirements than those presented in this study. While this holds, the adopted methodology was chosen to provide a focused analysis of the code synthesis capabilities of LLMs without over saturating the results with added complexity due to interdependencies. Thus, this reflects the quality expected in real-world software tasks. In addition, this study focused on the Python programing language, which has been well represented in the training data of many LLMs for code. However, many other popular programing languages may not be equally represented in the training data, which limits the conclusions to Python. In addition, evaluation focuses three primary nonfunctional requirements only. While this method is robust, it may not detect every nuance of code quality that a human reviewer may identify, potentially missing finer aspects of maintainability, reliability, and performance efficiency. Finally, this study did not consider the dynamic capabilities of LLMs, such as learning from feedback, which can enhance code quality over time.

The aforementioned limitations may also pose a risk to the validity and reproducibility of this research. The additional threats include LLM configuration and prompting. Previous research has documented the influence of varying prompts~\cite{mastropaolo2023robustness}, a similar insight was found in this research, thus suggesting that different settings or prompt structures could derive different results than those documented. In addition, the popular static code analysis tools used in this study depend on their accuracy and comprehensiveness. Although they were selected for their popularity, comprehensiveness, and ongoing active and contributing developer base, these tools may not comprehensively capture all relevant aspects of code quality or misinterpret code segments, leading to skewed results. Finally and arguably most significant is the rapidly evolving nature of technology, with frequent updates to LLMs. Considering the rate at which these LLMs are released, it could make the findings less relevant or accurate over time, necessitating ongoing updates to the research to maintain their applicability.

%%%%%%%%%%%%%%%%%%%%%%%%%%%%%%%%%%%%%%%%%%%%%%%%%%%%%%%%%%%%%%%%%%%%%%%%
\section{Conclusions and Future Work}
%%%%%%%%%%%%%%%%%%%%%%%%%%%%%%%%%%%%%%%%%%%%%%%%%%%%%%%%%%%%%%%%%%%%%%%%
This paper addresses the critical question of \textit{how LLMs can be effectively integrated into modern software development processes}. \textbf{P}rogrammatic \textbf{E}xcellence via \textbf{LL}M \textbf{I}teration (\textbf{PELLI}) framework was introduced. PELLI supports practitioners in effectively implementing LLMs within their workflows. At the heart of PELLI is the iteration and analysis of code changes, which promotes the harmonious integration of LLMs and human developers. This research and the introduction of PELLI contribute to a more nuanced understanding of how LLMs can be optimized for practical use, providing critical insights into their operational integration and the quality of their output.

Using a rigorous methodological framework grounded in practical application, this study investigated the capabilities of five state-of-the-art LLMs. To aid this understanding, LLM capabilities are examined across three application domains, and prompts of varying lengths and specific characteristics are used to reflect diverse contexts in which they are likely to be implemented. The findings demonstrate that when LLMs are appropriately prompted and configured, they can produce code that meets or even exceeds the human developer baseline in terms of maintainability, reliability, and performance. Among the tested LLMs, GPT-4T demonstrated marginally superior performance. 

In future work, it will be interesting to apply the above research methodology to another programing language. In addition, we include a unit test to test the intended functionality of the proposed algorithms, which can provide beneficial insights into the LLM’s ability to satisfy requirements. In addition, another nonfunctional requirement, i.e., security, can be added for more comprehensive evaluation.

\bibliographystyle{elsarticle-num}
\bibliography{ref_short}
\end{document}